\documentclass[12pt]{article}
\usepackage{amssymb}
\usepackage{amsmath}

\begin{document}

\begin{titlepage}
\title{\bf Lagrangian Dynamics on Clifford K\"{a}hler Manifolds}
\author{ Mehmet Tekkoyun \footnote{Corresponding author. E-mail address: tekkoyun@pau.edu.tr; Tel: +902582953616; Fax: +902582953593}\\
{\small Department of Mathematics, Pamukkale University,}\\
{\small 20070 Denizli, Turkey}}
\date{\today}
\maketitle

\begin{abstract}

In this study, Clifford K\"{a}hler analogue of
Lagrangian dynamics is introduced. Also, the some geometrical and physical
results over the obtained Clifford K\"{a}hler dynamical
systems are discussed.



\end{abstract}
\end{titlepage}

\section{Introduction}

It is well-known that modern differential geometry express explicitly the
dynamics of Lagrangians. Therefore we explain that if $M$ is an $m$%
-dimensional configuration manifold and $L:TM\rightarrow \mathbf{R}$\textbf{%
\ }is a regular Lagrangian function, then there is a unique vector field $%
\xi $ on $TM$ such that dynamics equations is determined by
\begin{equation}
i_{\xi }\Phi _{L}=dE_{L}  \label{1.1}
\end{equation}%
where $\Phi _{L}$ indicates the symplectic form. The triple $(TM,\Phi
_{L},\xi )$ is named \textit{Lagrangian system} on the tangent bundle $TM$ $%
. $

It is known, there are many studies about Lagrangian mechanics, formalisms,
systems and equations such that real, complex, paracomplex and other
analogues \cite{deleon, tekkoyun} and there in. So, it may be produced
different analogues in different spaces. The \ goal of finding new dynamics
equations is both a new expansion and contribution to science to explain
physical events.

Sir William Rowan Hamilton invented quaternions as an extension to the
complex numbers. Hamilton's defining relation is most succinctly written as:

\begin{equation}
i^{2}=j^{2}=k^{2}=ijk=-1  \label{1.2}
\end{equation}%
If it is compared to the calculus of vectors, quaternions have slipped into
the realm of obscurity. They do however still find use in the computation of
rotations. A lot of physical laws in classical, relativistic, and quantum
mechanics can be written pleasantly by means of quaternions. Some physicists
hope they will find deeper understanding of the universe by restating basic
principles in terms of quaternion algebra. It is well-known that quaternions
are useful for representing rotations in both quantum and classical
mechanics \cite{dan} . Clifford manifolds are also quaternion manifolds. One
may say that the above properties yield for Clifford manifold. It is seen
that Clifford manifold has also an important role in physical fields. For
example, considering Riemannian manifold ($M^{8n},g$) is the simplest
example of Clifford K\"{a}hler manifold, author obtained Euler-Lagrange
equations using a canonical local basis $\{J_{1},J_{2},J_{3}\}$ of $V$ on
the standard Clifford K\"{a}hler manifold $(\mathbf{R}^{8n},V)$ in \cite%
{tekkoyun1}.

In this study, Euler-Lagrange equations for Lagrangian systems on Clifford K%
\"{a}hler manifold have been introduced.

\section{Preliminaries}

In this paper, all mathematical objects and mappings are assumed to be
smooth, i.e. infinitely differentiable and Einstein convention of
summarizing is adopted. $\mathcal{F}(M)$, $\chi (M)$ and $\Lambda ^{1}(M)$
define the set of functions on $M$, the set of vector fields on $M$ and the
set of 1-forms on $M$, respectively.

\subsection{Clifford K\"{a}hler Manifolds}

Now, here we extend and rewrite the main concepts and structures given in
\cite{yano, burdujan} . Let $M$ be a real smooth manifold of dimension $m.$
Assume that there is a 6-dimensional vector bundle $V$ consisting of $%
F_{i}(i=1,2,...,6)$ tensors of type (1,1) over $M.$ Such a local basis $%
\{F_{1},F_{2},...,F_{6}\}$ is named a canonical local basis of the bundle $V$
in a neighborhood $U$ of $M$. Then $V$ is said to be an almost Clifford
structure in $M$. The pair $(M,V)$ is called an almost Clifford manifold
with $V$. Thus, an almost Clifford manifold $M$ is of dimension $m=8n.$ If
there exists on $(M,V)$ a global basis $\{F_{1},F_{2},...,F_{6}\},$ then $%
(M,V)$ is said to be an almost Clifford manifold; the basis $%
\{F_{1},F_{2},...,F_{6}\}$ is called a global basis for $V$.

An almost Clifford connection on the almost Clifford manifold $(M,V)$ is a
linear connection $\nabla $ on $M$ which preserves by parallel transport the
vector bundle $V$. This means that if $\Phi $ is a cross-section
(local-global) of the bundle $V$, then $\nabla _{X}\Phi $ is also a
cross-section (local-global, respectively) of $V$, $X$ being an arbitrary
vector field of $M$.

If for any canonical basis $\{J_{i}\},$ $i=\overline{1,6}$ of $V$ in a
coordinate neighborhood $U$, the identities
\begin{equation}
g(J_{i}X,J_{i}Y)=g(X,Y),\text{ }\forall X,Y\in \chi (M),\text{ }\
i=1,2,...,6,  \label{2.2}
\end{equation}%
hold, the triple $(M,g,V)$ is said to be an almost Clifford Hermitian
manifold or metric Clifford manifold denoting by $V$ an almost Clifford
structure $V$ and by $g$ a Riemannian metric and by $(g,V)$ an almost
Clifford metric structure$.$

Since each $J_{i}(i=1,2,...,6)$ is almost Hermitian structure\ with respect
to $g$, setting

\begin{equation}
\Phi _{i}(X,Y)=g(J_{i}X,Y),~\text{ }i=1,2,...,6,  \label{2.3}
\end{equation}

for any vector fields $X$ and $Y$, we see that $\Phi _{i}$ are 6 local
2-forms.

If the Levi-Civita connection $\nabla =\nabla ^{g}$ on $(M,g,V)$ preserves
the vector bundle $V$ by parallel transport, then $(M,g,V)$ is named a
Clifford K\"{a}hler manifold, and an almost Clifford structure $\Phi _{i}$
of $M$ is said to be a Clifford K\"{a}hler structure. Assume that let
\begin{equation*}
\left\{
x_{i},x_{n+i},x_{2n+i},x_{3n+i},x_{4n+i},x_{5n+i},x_{6n+i},x_{7n+i}\right\}
,i=\overline{1,n}
\end{equation*}%
be a real coordinate system on $(M,V).$ Then we determine by
\begin{eqnarray*}
&&\left\{ \frac{\partial }{\partial x_{i}},\frac{\partial }{\partial x_{n+i}}%
,\frac{\partial }{\partial x_{2n+i}},\frac{\partial }{\partial x_{3n+i}},%
\frac{\partial }{\partial x_{4n+i}},\frac{\partial }{\partial x_{5n+i}},%
\frac{\partial }{\partial x_{6n+i}},\frac{\partial }{\partial x_{7n+i}}%
\right\} , \\
&&%
\{dx_{i},dx_{n+i},dx_{2n+i},dx_{3n+i},dx_{4n+i},dx_{5n+i},dx_{6n+i},dx_{7n+i}\}
\end{eqnarray*}%
the natural bases over $\mathbf{R}$ of the tangent space $T(M)$ and the
cotangent space $T^{\ast }(M)$ of $M,$ respectively$.$ By structures $%
\{J_{i}\},$ $i=\overline{1,6}$ the following expressions are calculated

$%
\begin{array}{ccc}
J_{1}(\frac{\partial }{\partial x_{i}})=\frac{\partial }{\partial x_{n+i}} &
J_{2}(\frac{\partial }{\partial x_{i}})=\frac{\partial }{\partial x_{2n+i}}
& J_{3}(\frac{\partial }{\partial x_{i}})=\frac{\partial }{\partial x_{3n+i}}
\\
\text{ }J_{1}(\frac{\partial }{\partial x_{n+i}})=-\frac{\partial }{\partial
x_{i}} & J_{2}(\frac{\partial }{\partial x_{n+i}})=-\frac{\partial }{%
\partial x_{4n+i}} & J_{3}(\frac{\partial }{\partial x_{n+i}})=-\frac{%
\partial }{\partial x_{5n+i}} \\
J_{1}(\frac{\partial }{\partial x_{2n+i}})=\frac{\partial }{\partial x_{4n+i}%
} & J_{2}(\frac{\partial }{\partial x_{2n+i}})=-\frac{\partial }{\partial
x_{i}} & J_{3}(\frac{\partial }{\partial x_{2n+i}})=-\frac{\partial }{%
\partial x_{6n+i}} \\
J_{1}(\frac{\partial }{\partial x_{3n+i}})=\frac{\partial }{\partial x_{5n+i}%
} & J_{2}(\frac{\partial }{\partial x_{3n+i}})=\frac{\partial }{\partial
x_{6n+i}} & J_{3}(\frac{\partial }{\partial x_{3n+i}})=-\frac{\partial }{%
\partial x_{i}} \\
J_{1}(\frac{\partial }{\partial x_{4n+i}})=-\frac{\partial }{\partial
x_{2n+i}} & J_{2}(\frac{\partial }{\partial x_{4n+i}})=\frac{\partial }{%
\partial x_{n+i}} & J_{3}(\frac{\partial }{\partial x_{4n+i}})=\frac{%
\partial }{\partial x_{7n+i}} \\
J_{1}(\frac{\partial }{\partial x_{5n+i}})=-\frac{\partial }{\partial
x_{3n+i}} & J_{2}(\frac{\partial }{\partial x_{5n+i}})=-\frac{\partial }{%
\partial x_{7n+i}} & J_{3}(\frac{\partial }{\partial x_{5n+i}})=\frac{%
\partial }{\partial x_{n+i}} \\
\text{ }J_{1}(\frac{\partial }{\partial x_{6n+i}})=\frac{\partial }{\partial
x_{7n+i}} & J_{2}(\frac{\partial }{\partial x_{6n+i}})=-\frac{\partial }{%
\partial x_{3n+i}} & J_{3}(\frac{\partial }{\partial x_{6n+i}})=\frac{%
\partial }{\partial x_{2n+i}} \\
J_{1}(\frac{\partial }{\partial x_{7n+i}})=-\frac{\partial }{\partial
x_{6n+i}} & J_{2}(\frac{\partial }{\partial x_{7n+i}})=\frac{\partial }{%
\partial x_{5n+i}} & J_{3}(\frac{\partial }{\partial x_{7n+i}})=-\frac{%
\partial }{\partial x_{4n+i}}%
\end{array}%
$

$%
\begin{array}{ccc}
J_{4}(\frac{\partial }{\partial x_{i}})=\frac{\partial }{\partial x_{4n+i}}
& J_{5}(\frac{\partial }{\partial x_{i}})=\frac{\partial }{\partial x_{5n+i}}
& J_{6}(\frac{\partial }{\partial x_{i}})=\frac{\partial }{\partial x_{6n+i}}
\\
\text{ }J_{4}(\frac{\partial }{\partial x_{n+i}})=-\frac{\partial }{\partial
x_{2n+i}} & \text{ }J_{5}(\frac{\partial }{\partial x_{n+i}})=-\frac{%
\partial }{\partial x_{3n+i}} & J_{6}(\frac{\partial }{\partial x_{n+i}})=-%
\frac{\partial }{\partial x_{7n+i}} \\
J_{4}(\frac{\partial }{\partial x_{2n+i}})=\frac{\partial }{\partial x_{n+i}}
& J_{5}(\frac{\partial }{\partial x_{2n+i}})=-\frac{\partial }{\partial
x_{7n+i}} & J_{6}(\frac{\partial }{\partial x_{2n+i}})=-\frac{\partial }{%
\partial x_{3n+i}} \\
J_{4}(\frac{\partial }{\partial x_{3n+i}})=-\frac{\partial }{\partial
x_{7n+i}} & J_{5}(\frac{\partial }{\partial x_{3n+i}})=\frac{\partial }{%
\partial x_{n+i}} & J_{6}(\frac{\partial }{\partial x_{3n+i}})=\frac{%
\partial }{\partial x_{2n+i}} \\
J_{4}(\frac{\partial }{\partial x_{4n+i}})=-\frac{\partial }{\partial x_{i}}
& J_{5}(\frac{\partial }{\partial x_{4n+i}})=\frac{\partial }{\partial
x_{6n+i}} & J_{6}(\frac{\partial }{\partial x_{4n+i}})=\frac{\partial }{%
\partial x_{5n+i}} \\
\text{ }J_{4}(\frac{\partial }{\partial x_{5n+i}})=\frac{\partial }{\partial
x_{6n+i}} & J_{5}(\frac{\partial }{\partial x_{5n+i}})=-\frac{\partial }{%
\partial x_{i}} & J_{6}(\frac{\partial }{\partial x_{5n+i}})=-\frac{\partial
}{\partial x_{4n+i}} \\
J_{4}(\frac{\partial }{\partial x_{6n+i}})=-\frac{\partial }{\partial
x_{5n+i}} & J_{5}(\frac{\partial }{\partial x_{6n+i}})=-\frac{\partial }{%
\partial x_{4n+i}} & \text{ }J_{6}(\frac{\partial }{\partial x_{6n+i}})=-%
\frac{\partial }{\partial x_{i}} \\
J_{4}(\frac{\partial }{\partial x_{7n+i}})=\frac{\partial }{\partial x_{3n+i}%
} & J_{5}(\frac{\partial }{\partial x_{7n+i}})=\frac{\partial }{\partial
x_{2n+i}} & J_{6}(\frac{\partial }{\partial x_{7n+i}})=\frac{\partial }{%
\partial x_{n+i}}%
\end{array}%
$

\section{Lagrangian Mechanics}

In this section, we introduce Euler-Lagrange equations for quantum and
classical mechanics by means of a canonical local basis $\{J_{i}\},$ $i=%
\overline{1,6}$ of $V$ on Clifford K\"{a}hler manifold $(M,V).$ We say that
the Euler-Lagrange equations using basis $\{J_{1},J_{2},J_{3}\}$ of $V$ on $(%
\mathbf{R}^{8n},V)$ are introduced in \cite{tekkoyun1}$.$ In this study, we
obtain that they are the same as the equations \ obtained by operators $%
J_{1},J_{2},J_{3}$ of $V$ on Clifford K\"{a}hler manifold $(M,V).$ If we
reexpress them, they are respectively:

first:

$%
\begin{array}{c}
\frac{\partial }{\partial t}\left( \frac{\partial L}{\partial x_{i}}\right) +%
\frac{\partial L}{\partial x_{n+i}}=0,\frac{\partial }{\partial t}\left(
\frac{\partial L}{\partial x_{n+i}}\right) -\frac{\partial L}{\partial x_{i}}%
=0,\frac{\partial }{\partial t}\left( \frac{\partial L}{\partial x_{2n+i}}%
\right) +\frac{\partial L}{\partial x_{4n+i}}=0, \\
\frac{\partial }{\partial t}\left( \frac{\partial L}{\partial x_{3n+i}}%
\right) +\frac{\partial L}{\partial x_{5n+i}}=0,\frac{\partial }{\partial t}%
\left( \frac{\partial L}{\partial x_{4n+i}}\right) -\frac{\partial L}{%
\partial x_{2n+i}}=0,\frac{\partial }{\partial t}\left( \frac{\partial L}{%
\partial x_{5n+i}}\right) -\frac{\partial L}{\partial x_{3n+i}}=0, \\
\frac{\partial }{\partial t}\left( \frac{\partial L}{\partial x_{6n+i}}%
\right) +\frac{\partial L}{\partial x_{7n+i}}=0,\frac{\partial }{\partial t}%
\left( \frac{\partial L}{\partial x_{7n+i}}\right) -\frac{\partial L}{%
\partial x_{6n+i}}=0.%
\end{array}%
$

second:

$%
\begin{array}{c}
\frac{\partial }{\partial t}\left( \frac{\partial L}{\partial x_{i}}\right) +%
\frac{\partial L}{\partial x_{2n+i}}=0,\frac{\partial }{\partial t}\left(
\frac{\partial L}{\partial x_{n+i}}\right) -\frac{\partial L}{\partial
x_{4n+i}}=0,\frac{\partial }{\partial t}\left( \frac{\partial L}{\partial
x_{2n+i}}\right) -\frac{\partial L}{\partial x_{i}}=0, \\
\frac{\partial }{\partial t}\left( \frac{\partial L}{\partial x_{3n+i}}%
\right) +\frac{\partial L}{\partial x_{6n+i}}=0,\frac{\partial }{\partial t}%
\left( \frac{\partial L}{\partial x_{4n+i}}\right) +\frac{\partial L}{%
\partial x_{n+i}}=0,\frac{\partial }{\partial t}\left( \frac{\partial L}{%
\partial x_{5n+i}}\right) -\frac{\partial L}{\partial x_{7n+i}}=0, \\
\frac{\partial }{\partial t}\left( \frac{\partial L}{\partial x_{6n+i}}%
\right) -\frac{\partial L}{\partial x_{3n+i}}=0,\frac{\partial }{\partial t}%
\left( \frac{\partial L}{\partial x_{7n+i}}\right) +\frac{\partial L}{%
\partial x_{5n+i}}=0.%
\end{array}%
$

third:

$%
\begin{array}{c}
\frac{\partial }{\partial t}\left( \frac{\partial L}{\partial x_{i}}\right) +%
\frac{\partial L}{\partial x_{3n+i}}=0,\frac{\partial }{\partial t}\left(
\frac{\partial L}{\partial x_{n+i}}\right) -\frac{\partial L}{\partial
x_{5n+i}}=0,\frac{\partial }{\partial t}\left( \frac{\partial L}{\partial
x_{2n+i}}\right) -\frac{\partial L}{\partial x_{6n+i}}=0, \\
\frac{\partial }{\partial t}\left( \frac{\partial L}{\partial x_{3n+i}}%
\right) -\frac{\partial L}{\partial x_{i}}=0,\frac{\partial }{\partial t}%
\left( \frac{\partial L}{\partial x_{4n+i}}\right) +\frac{\partial L}{%
\partial x_{7n+i}}=0,\frac{\partial }{\partial t}\left( \frac{\partial L}{%
\partial x_{5n+i}}\right) +\frac{\partial L}{\partial x_{n+i}}=0, \\
\frac{\partial }{\partial t}\left( \frac{\partial L}{\partial x_{6n+i}}%
\right) +\frac{\partial L}{\partial x_{2n+i}}=0,\frac{\partial }{\partial t}%
\left( \frac{\partial L}{\partial x_{7n+i}}\right) -\frac{\partial L}{%
\partial x_{4n+i}}=0.%
\end{array}%
$

Here, only we derive Euler-Lagrange equations using operators $%
J_{4},J_{5},J_{6}$ of $V$ on Clifford K\"{a}hler manifold $(M,V).$

Fourth, let $J_{4}$ take a local basis component on Clifford K\"{a}hler
manifold $(M,V),$ and $\left\{
x_{i},x_{n+i},x_{2n+i},x_{3n+i},x_{4n+i},x_{5n+i},x_{6n+i},x_{7n+i}\right\}
, $ $i=\overline{1,n}$ be its coordinate functions. Let semispray be the
vector field $\xi $ defined by%
\begin{equation}
\begin{array}{c}
\xi =X^{i}\frac{\partial }{\partial x_{i}}+X^{n+i}\frac{\partial }{\partial
x_{n+i}}+X^{2n+i}\frac{\partial }{\partial x_{2n+i}}+X^{3n+i}\frac{\partial
}{\partial x_{3n+i}} \\
+X^{4n+i}\frac{\partial }{\partial x_{4n+i}}+X^{5n+i}\frac{\partial }{%
\partial x_{5n+i}}+X^{6n+i}\frac{\partial }{\partial x_{6n+i}}+X^{7n+i}\frac{%
\partial }{\partial x_{7n+i}},%
\end{array}
\label{3.1}
\end{equation}%
where%
\begin{eqnarray*}
X^{i} &=&\overset{.}{x_{i}},X^{n+i}=\overset{.}{x}_{n+i},X^{2n+i}=\overset{.}%
{x}_{2n+i},X^{3n+i}=\overset{.}{x}_{3n+i}, \\
X^{4n+i} &=&\overset{.}{x_{4n+i}},X^{5n+i}=\overset{.}{x}_{5n+i},X^{6n+i}=%
\overset{.}{x}_{6n+i},X^{7n+i}=\overset{.}{x}_{7n+i}
\end{eqnarray*}

and the dot indicates the derivative with respect to time $t$. The vector
fields determined by%
\begin{equation}
\begin{array}{c}
V_{J_{4}}=J_{4}(\xi )=X^{i}\frac{\partial }{\partial x_{4n+i}}-X^{n+i}\frac{%
\partial }{\partial x_{2n+i}}+X^{2n+i}\frac{\partial }{\partial x_{n+i}}%
-X^{3n+i}\frac{\partial }{\partial x_{7n+i}} \\
-X^{4n+i}\frac{\partial }{\partial x_{i}}+X^{5n+i}\frac{\partial }{\partial
x_{6n+i}}-X^{6n+i}\frac{\partial }{\partial x_{5n+i}}+X^{7n+i}\frac{\partial
}{\partial x_{3n+i}},%
\end{array}
\label{3.2}
\end{equation}%
is named \textit{Liouville vector field} on Clifford K\"{a}hler manifold $%
(M,V)$. The maps explained by $T,P:M\rightarrow \mathbf{R}$ such that%
\begin{equation*}
T=\frac{1}{2}m_{i}(\overset{.}{x_{i}}^{2}+\overset{.}{x}%
_{n+i}^{2}+x_{2n+i}^{2}+\overset{.}{x}_{3n+i}^{2}+\overset{.}{x_{4n+i}}^{2}+%
\overset{.}{x}_{5n+i}^{2}+x_{6n+i}^{2}+\overset{.}{x}_{7n+i}^{2}),\text{ \ }%
P=m_{i}gh
\end{equation*}

are said to be \textit{the kinetic energy} and \textit{the potential energy
of the system,} respectively.\textit{\ }Here\textit{\ }$m_{i},g$ and $h$
stand for mass of a mechanical system having $m$ particles, the gravity
acceleration and distance to the origin of a mechanical system on Clifford K%
\"{a}hler manifold $(M,V)$, respectively. Then $L:M\rightarrow R$ is a map
that satisfies the conditions; i) $L=T-P$ is a \textit{Lagrangian function,
ii)} the function given by $E_{L}^{J_{4}}=V_{J_{4}}(L)-L,$ is\textit{\
energy function}.

The operator $i_{J_{4}}$ induced by $J_{4}$ and defined by%
\begin{equation}
i_{J_{4}}\omega (X_{1},X_{2},...,X_{r})=\sum_{i=1}^{r}\omega
(X_{1},...,J_{4}X_{i},...,X_{r}),  \label{3.3}
\end{equation}

is called \textit{vertical derivation, }where $\omega \in \wedge ^{r}M,$ $%
X_{i}\in \chi (M).$ The \textit{vertical differentiation} $d_{J_{4}}$ is
determined by%
\begin{equation}
d_{J_{4}}=[i_{J_{4}},d]=i_{J_{4}}d-di_{J_{4}}  \label{3.4}
\end{equation}%
where $d$ is the usual exterior derivation. We saw that the closed Clifford K%
\"{a}hler form is the closed 2-form given by $\Phi
_{L}^{J_{4}}=-dd_{_{J_{4}}}L$ such that%
\begin{eqnarray*}
d_{_{J_{4}}} &=&\frac{\partial }{\partial x_{4n+i}}dx_{i}-\frac{\partial }{%
\partial x_{2n+i}}dx_{n+i}+\frac{\partial }{\partial x_{n+i}}dx_{2n+i}-\frac{%
\partial }{\partial x_{7n+i}}dx_{3n+i} \\
&&-\frac{\partial }{\partial x_{i}}dx_{4n+i}+\frac{\partial }{\partial
x_{6n+i}}dx_{5n+i}-\frac{\partial }{\partial x_{5n+i}}dx_{6n+i}+\frac{%
\partial }{\partial x_{3n+i}}dx_{7n+i}
\end{eqnarray*}

determined by operator%
\begin{equation}
d_{_{J_{4}}}:\mathcal{F}(M)\rightarrow \wedge ^{1}{}M.  \label{3.5}
\end{equation}

Then

$\Phi _{L}^{J_{4}}=-\frac{\partial ^{2}L}{\partial x_{j}\partial x_{4n+i}}%
dx_{j}\wedge dx_{i}+\frac{\partial ^{2}L}{\partial x_{j}\partial x_{2n+i}}%
dx_{j}\wedge dx_{n+i}-\frac{\partial ^{2}L}{\partial x_{j}\partial x_{n+i}}%
dx_{j}\wedge dx_{2n+i}$

$+\frac{\partial ^{2}L}{\partial x_{j}\partial x_{7n+i}}dx_{j}\wedge
dx_{3n+i}+\frac{\partial ^{2}L}{\partial x_{j}\partial x_{i}}dx_{j}\wedge
dx_{4n+i}-\frac{\partial ^{2}L}{\partial x_{j}\partial x_{6n+i}}dx_{j}\wedge
dx_{5n+i}$

$+\frac{\partial ^{2}L}{\partial x_{j}\partial x_{5n+i}}dx_{j}\wedge
dx_{6n+i}-\frac{\partial ^{2}L}{\partial x_{j}\partial x_{3n+i}}dx_{j}\wedge
dx_{7n+i}-\frac{\partial ^{2}L}{\partial x_{n+j}\partial x_{4n+i}}%
dx_{n+j}\wedge dx_{i}$

$+\frac{\partial ^{2}L}{\partial x_{n+j}\partial x_{2n+i}}dx_{n+j}\wedge
dx_{n+i}-\frac{\partial ^{2}L}{\partial x_{n+j}\partial x_{n+i}}%
dx_{n+j}\wedge dx_{2n+i}+\frac{\partial ^{2}L}{\partial x_{n+j}\partial
x_{7n+i}}dx_{n+j}\wedge dx_{3n+i}$

$+\frac{\partial ^{2}L}{\partial x_{n+j}\partial x_{i}}dx_{n+j}\wedge
dx_{4n+i}-\frac{\partial ^{2}L}{\partial x_{n+j}\partial x_{6n+i}}%
dx_{n+j}\wedge dx_{5n+i}+\frac{\partial ^{2}L}{\partial x_{n+j}\partial
x_{5n+i}}dx_{n+j}\wedge dx_{6n+i}$

$-\frac{\partial ^{2}L}{\partial x_{n+j}\partial x_{7n+i}}dx_{n+j}\wedge
dx_{7n+i}-\ \frac{\partial ^{2}L}{\partial x_{2n+j}\partial x_{4n+i}}%
dx_{2n+j}\wedge dx_{i}+\frac{\partial ^{2}L}{\partial x_{2n+j}\partial
x_{2n+i}}dx_{2n+j}\wedge dx_{n+i}$

$-\frac{\partial ^{2}L}{\partial x_{2n+j}\partial x_{n+i}}dx_{2n+j}\wedge
dx_{2n+i}+\frac{\partial ^{2}L}{\partial x_{2n+j}\partial x_{7n+i}}%
dx_{2n+j}\wedge dx_{3n+i}+\frac{\partial ^{2}L}{\partial x_{2n+j}\partial
x_{i}}dx_{2n+j}\wedge dx_{4n+i}$

$-\frac{\partial ^{2}L}{\partial x_{2n+j}\partial x_{6n+i}}dx_{2n+j}\wedge
dx_{5n+i}+\frac{\partial ^{2}L}{\partial x_{2n+j}\partial x_{5n+i}}%
dx_{2n+j}\wedge dx_{6n+i}-\frac{\partial ^{2}L}{\partial x_{2n+j}\partial
x_{7n+i}}dx_{2n+j}\wedge dx_{7n+i}$

$-\frac{\partial ^{2}L}{\partial x_{3n+j}\partial x_{4n+i}}dx_{3n+j}\wedge
dx_{i}+\frac{\partial ^{2}L}{\partial x_{3n+j}\partial x_{2n+i}}%
dx_{3n+j}\wedge dx_{n+i}-\frac{\partial ^{2}L}{\partial x_{3n+j}\partial
x_{n+i}}dx_{3n+j}\wedge dx_{2n+i}$

$+\frac{\partial ^{2}L}{\partial x_{3n+j}\partial x_{7n+i}}dx_{3n+j}\wedge
dx_{3n+i}+\frac{\partial ^{2}L}{\partial x_{3n+j}\partial x_{i}}%
dx_{3n+j}\wedge dx_{4n+i}-\frac{\partial ^{2}L}{\partial x_{3n+j}\partial
x_{6n+i}}dx_{3n+j}\wedge dx_{5n+i}$

$+\frac{\partial ^{2}L}{\partial x_{3n+j}\partial x_{5n+i}}dx_{3n+j}\wedge
dx_{6n+i}-\frac{\partial ^{2}L}{\partial x_{3n+j}\partial x_{7n+i}}%
dx_{3n+j}\wedge dx_{7n+i}-\frac{\partial L}{\partial x_{4n+j}\partial
x_{4n+i}}dx_{4n+j}\wedge dx_{i}$

$+\frac{\partial L}{\partial x_{4n+j}\partial x_{2n+i}}dx_{4n+j}\wedge
dx_{n+i}-\frac{\partial L}{\partial x_{4n+j}\partial x_{n+i}}dx_{4n+j}\wedge
dx_{2n+i}+\frac{\partial L}{\partial x_{4n+j}\partial x_{7n+i}}%
dx_{4n+j}\wedge dx_{3n+i}$

$+\frac{\partial ^{2}L}{\partial x_{4n+j}\partial x_{i}}dx_{4n+j}\wedge
dx_{4n+i}-\frac{\partial ^{2}L}{\partial x_{4n+j}\partial x_{6n+i}}%
dx_{4n+j}\wedge dx_{5n+i}+\frac{\partial ^{2}L}{\partial x_{4n+j}\partial
x_{5n+i}}dx_{4n+j}\wedge dx_{6n+i}$

$-\frac{\partial ^{2}L}{\partial x_{4n+j}\partial x_{3n+i}}dx_{4n+j}\wedge
dx_{7n+i}-\frac{\partial ^{2}L}{\partial x_{5n+j}\partial x_{4n+i}}%
dx_{5n+j}\wedge dx_{i}+\frac{\partial ^{2}L}{\partial x_{5n+j}\partial
x_{2n+i}}dx_{5n+j}\wedge dx_{n+i}$

$-\frac{\partial ^{2}L}{\partial x_{5n+j}\partial x_{n+i}}dx_{5n+j}\wedge
dx_{2n+i}+\frac{\partial ^{2}L}{\partial x_{5n+j}\partial x_{7n+i}}%
dx_{5n+j}\wedge dx_{3n+i}+\frac{\partial ^{2}L}{\partial x_{5n+j}\partial
x_{i}}dx_{5n+j}\wedge dx_{4n+i}$

$-\frac{\partial ^{2}L}{\partial x_{5n+j}\partial x_{6n+i}}dx_{5n+j}\wedge
dx_{5n+i}+\frac{\partial ^{2}L}{\partial x_{5n+j}\partial x_{5n+i}}%
dx_{5n+j}\wedge dx_{6n+i}-\frac{\partial ^{2}L}{\partial x_{5n+j}\partial
x_{3n+i}}dx_{5n+j}\wedge dx_{7n+i}$

$-\frac{\partial ^{2}L}{\partial x_{6n+j}\partial x_{4n+i}}dx_{6n+j}\wedge
dx_{i}+\frac{\partial ^{2}L}{\partial x_{6n+j}\partial x_{2n+i}}%
dx_{6n+j}\wedge dx_{n+i}-\frac{\partial ^{2}L}{\partial x_{6n+j}\partial
x_{n+i}}dx_{6n+j}\wedge dx_{2n+i}$

$+\frac{\partial ^{2}L}{\partial x_{6n+j}\partial x_{7n+i}}dx_{6n+j}\wedge
dx_{3n+i}+\frac{\partial ^{2}L}{\partial x_{6n+j}\partial x_{i}}%
dx_{6n+j}\wedge dx_{4n+i}-\frac{\partial ^{2}L}{\partial x_{6n+j}\partial
x_{6n+i}}dx_{6n+j}\wedge dx_{5n+i}$

$+\frac{\partial ^{2}L}{\partial x_{6n+j}\partial x_{5n+i}}dx_{6n+j}\wedge
dx_{6n+i}-\frac{\partial ^{2}L}{\partial x_{6n+j}\partial x_{7n+i}}%
dx_{6n+j}\wedge dx_{7n+i}-\frac{\partial ^{2}L}{\partial x_{7n+j}\partial
x_{4n+i}}dx_{7n+j}\wedge dx_{i}$

$+\frac{\partial ^{2}L}{\partial x_{7n+j}\partial x_{2n+i}}dx_{7n+j}\wedge
dx_{n+i}-\frac{\partial ^{2}L}{\partial x_{7n+j}\partial x_{n+i}}%
dx_{7n+j}\wedge dx_{2n+i}+\frac{\partial ^{2}L}{\partial x_{7n+j}\partial
x_{7n+i}}dx_{7n+j}\wedge dx_{3n+i}$

$+\frac{\partial ^{2}L}{\partial x_{7n+j}\partial x_{i}}dx_{7n+j}\wedge
dx_{4n+i}-\frac{\partial ^{2}L}{\partial x_{7n+j}\partial x_{6n+i}}%
dx_{7n+j}\wedge dx_{5n+i}+\frac{\partial ^{2}L}{\partial x_{7n+j}\partial
x_{5n+i}}dx_{7n+j}\wedge dx_{6n+i}$

$-\frac{\partial ^{2}L}{\partial x_{7n+j}\partial x_{3n+i}}dx_{7n+j}\wedge
dx_{7n+i}$

Let $\xi $ be the second order differential equation by determined \textbf{%
Eq. }(\ref{1.1}) and given by \textbf{Eq. }(\ref{3.1}) and

$i_{\xi }\Phi _{L}^{J_{1}}=-X^{i}\frac{\partial ^{2}L}{\partial
x_{j}\partial x_{4n+i}}\delta _{i}^{j}dx_{i}+X^{i}\frac{\partial ^{2}L}{%
\partial x_{j}\partial x_{4n+i}}dx_{j}+X^{i}\frac{\partial ^{2}L}{\partial
x_{j}\partial x_{2n+i}}\delta _{i}^{j}dx_{n+i}-X^{n+i}\frac{\partial ^{2}L}{%
\partial x_{j}\partial x_{2n+i}}dx_{j}$

$-X^{i}\frac{\partial ^{2}L}{\partial x_{j}\partial x_{n+i}}\delta
_{i}^{j}dx_{2n+i}+X^{2n+i}\frac{\partial ^{2}L}{\partial x_{j}\partial
x_{n+i}}dx_{j}+X^{i}\frac{\partial ^{2}L}{\partial x_{j}\partial x_{7n+i}}%
\delta _{i}^{j}dx_{3n+i}-X^{3n+i}\frac{\partial ^{2}L}{\partial
x_{j}\partial x_{7n+i}}dx_{j}$

$+X^{i}\frac{\partial ^{2}L}{\partial x_{j}\partial x_{i}}\delta
_{i}^{j}dx_{4n+i}-X^{4n+i}\frac{\partial ^{2}L}{\partial x_{j}\partial x_{i}}%
dx_{j}-X^{i}\frac{\partial ^{2}L}{\partial x_{j}\partial x_{6n+i}}\delta
_{i}^{j}dx_{5n+i}+X^{5n+i}\frac{\partial ^{2}L}{\partial x_{j}\partial
x_{6n+i}}dx_{j}$

$+X^{i}\frac{\partial ^{2}L}{\partial x_{j}\partial x_{5n+i}}\delta
_{i}^{j}dx_{6n+i}-X^{6n+i}\frac{\partial ^{2}L}{\partial x_{j}\partial
x_{5n+i}}dx_{j}-X^{i}\frac{\partial ^{2}L}{\partial x_{j}\partial x_{3n+i}}%
\delta _{i}^{j}dx_{7n+i}+X^{7n+i}\frac{\partial ^{2}L}{\partial
x_{j}\partial x_{3n+i}}dx_{j}$

$-X^{n+i}\frac{\partial ^{2}L}{\partial x_{n+j}\partial x_{4n+i}}\delta
_{n+i}^{n+j}dx_{i}+X^{n+i}\frac{\partial ^{2}L}{\partial x_{n+j}\partial
x_{4n+i}}dx_{n+j}+X^{n+i}\frac{\partial ^{2}L}{\partial x_{n+j}\partial
x_{2n+i}}\delta _{n+i}^{n+j}dx_{n+i}$

$-X^{n+i}\frac{\partial ^{2}L}{\partial x_{n+j}\partial x_{2n+i}}%
dx_{n+j}-X^{n+i}\frac{\partial ^{2}L}{\partial x_{n+j}\partial x_{n+i}}%
\delta _{n+i}^{n+j}dx_{2n+i}\ +X^{2n+i}\frac{\partial ^{2}L}{\partial
x_{n+j}\partial x_{n+i}}dx_{n+j}$

$+X^{n+i}\frac{\partial ^{2}L}{\partial x_{n+j}\partial x_{7n+i}}\delta
_{n+i}^{n+j}dx_{3n+i}-X^{3n+i}\frac{\partial ^{2}L}{\partial x_{n+j}\partial
x_{7n+i}}dx_{n+j}+X^{n+i}\frac{\partial ^{2}L}{\partial x_{n+j}\partial x_{i}%
}\delta _{n+i}^{n+j}dx_{4n+i}$

$-X^{4n+i}\frac{\partial ^{2}L}{\partial x_{n+j}\partial x_{i}}%
dx_{n+j}-X^{n+i}\frac{\partial ^{2}L}{\partial x_{n+j}\partial x_{6n+i}}%
\delta _{n+i}^{n+j}dx_{5n+i}+X^{5n+i}\frac{\partial ^{2}L}{\partial
x_{n+j}\partial x_{6n+i}}dx_{n+j}$

$+X^{n+i}\frac{\partial ^{2}L}{\partial x_{n+j}\partial x_{5n+i}}\delta
_{n+i}^{n+j}dx_{6n+i}-X^{6n+i}\frac{\partial ^{2}L}{\partial x_{n+j}\partial
x_{5n+i}}dx_{n+j}-X^{n+i}\frac{\partial ^{2}L}{\partial x_{n+j}\partial
x_{7n+i}}\delta _{n+i}^{n+j}dx_{7n+i}$

$+X^{7n+i}\frac{\partial ^{2}L}{\partial x_{n+j}\partial x_{7n+i}}dx_{n+j}-\
X^{2n+i}\frac{\partial ^{2}L}{\partial x_{2n+j}\partial x_{4n+i}}\delta
_{2n+i}^{2n+j}dx_{i}+\ X^{i}\frac{\partial ^{2}L}{\partial x_{2n+j}\partial
x_{4n+i}}dx_{2n+j}$

$+X^{2n+i}\frac{\partial ^{2}L}{\partial x_{2n+j}\partial x_{2n+i}}\delta
_{2n+i}^{2n+j}dx_{n+i}-X^{n+i}\frac{\partial ^{2}L}{\partial
x_{2n+j}\partial x_{2n+i}}dx_{2n+j}-X^{2n+i}\frac{\partial ^{2}L}{\partial
x_{2n+j}\partial x_{n+i}}\delta _{2n+i}^{2n+j}dx_{2n+i}$

$+X^{2n+i}\frac{\partial ^{2}L}{\partial x_{2n+j}\partial x_{n+i}}%
dx_{2n+j}+X^{2n+i}\frac{\partial ^{2}L}{\partial x_{2n+j}\partial x_{7n+i}}%
\delta _{2n+i}^{2n+j}dx_{3n+i}-X^{3n+i}\frac{\partial ^{2}L}{\partial
x_{2n+j}\partial x_{7n+i}}dx_{2n+j}$

$+X^{2n+i}\frac{\partial ^{2}L}{\partial x_{2n+j}\partial x_{i}}\delta
_{2n+i}^{2n+j}dx_{4n+i}-X^{4n+i}\frac{\partial ^{2}L}{\partial
x_{2n+j}\partial x_{i}}dx_{2n+j}-X^{2n+i}\frac{\partial ^{2}L}{\partial
x_{2n+j}\partial x_{6n+i}}\delta _{2n+i}^{2n+j}dx_{5n+i}$

$+X^{5n+i}\frac{\partial ^{2}L}{\partial x_{2n+j}\partial x_{6n+i}}%
dx_{2n+j}+X^{2n+i}\frac{\partial ^{2}L}{\partial x_{2n+j}\partial x_{5n+i}}%
\delta _{2n+i}^{2n+j}dx_{6n+i}-X^{6n+i}\frac{\partial ^{2}L}{\partial
x_{2n+j}\partial x_{5n+i}}dx_{2n+j}$

$-X^{2n+i}\frac{\partial ^{2}L}{\partial x_{2n+j}\partial x_{7n+i}}\delta
_{2n+i}^{2n+j}dx_{7n+i}+X^{7n+i}\frac{\partial ^{2}L}{\partial
x_{2n+j}\partial x_{7n+i}}dx_{2n+j}-X^{3n+i}\frac{\partial ^{2}L}{\partial
x_{3n+j}\partial x_{4n+i}}\delta _{3n+i}^{3n+j}dx_{i}$

$+X^{i}\frac{\partial ^{2}L}{\partial x_{3n+j}\partial x_{4n+i}}%
dx_{3n+j}+X^{3n+i}\frac{\partial ^{2}L}{\partial x_{3n+j}\partial x_{2n+i}}%
\delta _{3n+i}^{3n+j}dx_{n+i}-X^{n+i}\frac{\partial ^{2}L}{\partial
x_{3n+j}\partial x_{2n+i}}dx_{3n+j}$

$-X^{3n+i}\frac{\partial ^{2}L}{\partial x_{3n+j}\partial x_{n+i}}\delta
_{3n+i}^{3n+j}dx_{2n+i}+X^{2n+i}\frac{\partial ^{2}L}{\partial
x_{3n+j}\partial x_{n+i}}dx_{3n+j}+X^{3n+i}\frac{\partial ^{2}L}{\partial
x_{3n+j}\partial x_{7n+i}}\delta _{3n+i}^{3n+j}dx_{3n+i}$

$-X^{3n+i}\frac{\partial ^{2}L}{\partial x_{3n+j}\partial x_{7n+i}}%
dx_{3n+j}+X^{3n+i}\frac{\partial ^{2}L}{\partial x_{3n+j}\partial x_{i}}%
\delta _{3n+i}^{3n+j}dx_{4n+i}-X^{4n+i}\frac{\partial ^{2}L}{\partial
x_{3n+j}\partial x_{i}}dx_{3n+j}$

$-X^{3n+i}\frac{\partial ^{2}L}{\partial x_{3n+j}\partial x_{6n+i}}\delta
_{3n+i}^{3n+j}dx_{5n+i}+X^{5n+i}\frac{\partial ^{2}L}{\partial
x_{3n+j}\partial x_{6n+i}}dx_{3n+j}+X^{3n+i}\frac{\partial ^{2}L}{\partial
x_{3n+j}\partial x_{5n+i}}\delta _{3n+i}^{3n+j}dx_{6n+i}$

$-X^{6n+i}\frac{\partial ^{2}L}{\partial x_{3n+j}\partial x_{5n+i}}%
dx_{3n+j}-X^{3n+i}\frac{\partial ^{2}L}{\partial x_{3n+j}\partial x_{7n+i}}%
\delta _{3n+i}^{3n+j}dx_{7n+i}+X^{7n+i}\frac{\partial ^{2}L}{\partial
x_{3n+j}\partial x_{7n+i}}dx_{3n+j}$

$-X^{4n+i}\frac{\partial L}{\partial x_{4n+j}\partial x_{4n+i}}\delta
_{4n+i}^{4n+j}dx_{i}+X^{i}\frac{\partial L}{\partial x_{4n+j}\partial
x_{4n+i}}dx_{4n+j}+X^{4n+i}\frac{\partial L}{\partial x_{4n+j}\partial
x_{2n+i}}\delta _{4n+i}^{4n+j}dx_{n+i}$

$-X^{n+i}\frac{\partial L}{\partial x_{4n+j}\partial x_{2n+i}}%
dx_{4n+j}-X^{4n+i}\frac{\partial L}{\partial x_{4n+j}\partial x_{n+i}}\delta
_{4n+i}^{4n+j}dx_{2n+i}\ \ +X^{2n+i}\frac{\partial L}{\partial
x_{4n+j}\partial x_{n+i}}dx_{4n+j}$

$+X^{4n+i}\frac{\partial L}{\partial x_{4n+j}\partial x_{7n+i}}\delta
_{4n+i}^{4n+j}dx_{3n+i}-X^{3n+i}\frac{\partial L}{\partial x_{4n+j}\partial
x_{7n+i}}dx_{4n+j}+X^{4n+i}\frac{\partial ^{2}L}{\partial x_{4n+j}\partial
x_{i}}\delta _{4n+i}^{4n+j}dx_{4n+i}$

$-X^{4n+i}\frac{\partial ^{2}L}{\partial x_{4n+j}\partial x_{i}}%
dx_{4n+j}-X^{4n+i}\frac{\partial ^{2}L}{\partial x_{4n+j}\partial x_{6n+i}}%
\delta _{4n+i}^{4n+j}dx_{5n+i}+X^{5n+i}\frac{\partial ^{2}L}{\partial
x_{4n+j}\partial x_{6n+i}}dx_{4n+j}$

$+X^{4n+i}\frac{\partial ^{2}L}{\partial x_{4n+j}\partial x_{5n+i}}\delta
_{4n+i}^{4n+j}dx_{6n+i}-X^{6n+i}\frac{\partial ^{2}L}{\partial
x_{4n+j}\partial x_{5n+i}}dx_{4n+j}-X^{4n+i}\frac{\partial ^{2}L}{\partial
x_{4n+j}\partial x_{3n+i}}\delta _{4n+i}^{4n+j}dx_{7n+i}$

$+X^{7n+i}\frac{\partial ^{2}L}{\partial x_{4n+j}\partial x_{3n+i}}%
dx_{4n+j}-X^{5n+i}\frac{\partial ^{2}L}{\partial x_{5n+j}\partial x_{4n+i}}%
\delta _{5n+i}^{5n+j}dx_{i}\ +X^{i}\frac{\partial ^{2}L}{\partial
x_{5n+j}\partial x_{4n+i}}dx_{5n+j}$

$+X^{5n+i}\frac{\partial ^{2}L}{\partial x_{5n+j}\partial x_{2n+i}}\delta
_{5n+i}^{5n+j}dx_{n+i}-X^{n+i}\frac{\partial ^{2}L}{\partial
x_{5n+j}\partial x_{2n+i}}dx_{5n+j}-X^{5n+i}\frac{\partial ^{2}L}{\partial
x_{5n+j}\partial x_{n+i}}\delta _{5n+i}^{5n+j}dx_{2n+i}$

$+X^{2n+i}\frac{\partial ^{2}L}{\partial x_{5n+j}\partial x_{n+i}}%
dx_{5n+j}+X^{5n+i}\frac{\partial ^{2}L}{\partial x_{5n+j}\partial x_{7n+i}}%
\delta _{5n+i}^{5n+j}dx_{3n+i}-X^{3n+i}\frac{\partial ^{2}L}{\partial
x_{5n+j}\partial x_{7n+i}}dx_{5n+j}$

$+X^{5n+i}\frac{\partial ^{2}L}{\partial x_{5n+j}\partial x_{i}}\delta
_{5n+i}^{5n+j}dx_{4n+i}-X^{4n+i}\frac{\partial ^{2}L}{\partial
x_{5n+j}\partial x_{i}}dx_{5n+j}-X^{5n+i}\frac{\partial ^{2}L}{\partial
x_{5n+j}\partial x_{6n+i}}\delta _{5n+i}^{5n+j}dx_{5n+i}$

$+X^{5n+i}\frac{\partial ^{2}L}{\partial x_{5n+j}\partial x_{6n+i}}%
dx_{5n+j}+X^{5n+i}\frac{\partial ^{2}L}{\partial x_{5n+j}\partial x_{5n+i}}%
\delta _{5n+i}^{5n+j}dx_{6n+i}-X^{6n+i}\frac{\partial ^{2}L}{\partial
x_{5n+j}\partial x_{5n+i}}dx_{5n+j}$

$-X^{5n+i}\frac{\partial ^{2}L}{\partial x_{5n+j}\partial x_{3n+i}}\delta
_{5n+i}^{5n+j}dx_{7n+i}+X^{7n+i}\frac{\partial ^{2}L}{\partial
x_{5n+j}\partial x_{3n+i}}dx_{5n+j}-X^{6n+i}\frac{\partial ^{2}L}{\partial
x_{6n+j}\partial x_{4n+i}}\delta _{6n+i}^{6n+j}dx_{i}$

$+X^{i}\frac{\partial ^{2}L}{\partial x_{6n+j}\partial x_{4n+i}}%
dx_{6n+j}+X^{6n+i}\frac{\partial ^{2}L}{\partial x_{6n+j}\partial x_{2n+i}}%
\delta _{6n+i}^{6n+j}dx_{n+i}-X^{n+i}\frac{\partial ^{2}L}{\partial
x_{6n+j}\partial x_{2n+i}}dx_{6n+j}$

$-X^{6n+i}\frac{\partial ^{2}L}{\partial x_{6n+j}\partial x_{n+i}}\delta
_{6n+i}^{6n+j}dx_{2n+i}+X^{2n+i}\frac{\partial ^{2}L}{\partial
x_{6n+j}\partial x_{n+i}}dx_{6n+j}+X^{6n+i}\frac{\partial ^{2}L}{\partial
x_{6n+j}\partial x_{7n+i}}\delta _{6n+i}^{6n+j}dx_{3n+i}$

$-X^{3n+i}\frac{\partial ^{2}L}{\partial x_{6n+j}\partial x_{7n+i}}%
dx_{6n+j}+X^{6n+i}\frac{\partial ^{2}L}{\partial x_{6n+j}\partial x_{i}}%
\delta _{6n+i}^{6n+j}dx_{4n+i}\ -X^{4n+i}\frac{\partial ^{2}L}{\partial
x_{6n+j}\partial x_{i}}dx_{6n+j}$

$-X^{6n+i}\frac{\partial ^{2}L}{\partial x_{6n+j}\partial x_{6n+i}}\delta
_{6n+i}^{6n+j}dx_{5n+i}+X^{5n+i}\frac{\partial ^{2}L}{\partial
x_{6n+j}\partial x_{6n+i}}dx_{6n+j}+X^{6n+i}\frac{\partial ^{2}L}{\partial
x_{6n+j}\partial x_{5n+i}}\delta _{6n+i}^{6n+j}dx_{6n+i}$

$-X^{6n+i}\frac{\partial ^{2}L}{\partial x_{6n+j}\partial x_{5n+i}}%
dx_{6n+j}-X^{6n+i}\frac{\partial ^{2}L}{\partial x_{6n+j}\partial x_{7n+i}}%
\delta _{6n+i}^{6n+j}dx_{7n+i}+X^{7n+i}\frac{\partial ^{2}L}{\partial
x_{6n+j}\partial x_{7n+i}}dx_{6n+j}$

$-X^{7n+i}\frac{\partial ^{2}L}{\partial x_{7n+j}\partial x_{4n+i}}\delta
_{7n+i}^{7n+j}dx_{i}+X^{i}\frac{\partial ^{2}L}{\partial x_{7n+j}\partial
x_{4n+i}}dx_{7n+j}+X^{7n+i}\frac{\partial ^{2}L}{\partial x_{7n+j}\partial
x_{2n+i}}\delta _{7n+i}^{7n+j}dx_{n+i}$

$-X^{n+i}\frac{\partial ^{2}L}{\partial x_{7n+j}\partial x_{2n+i}}%
dx_{7n+j}-X^{7n+i}\frac{\partial ^{2}L}{\partial x_{7n+j}\partial x_{n+i}}%
\delta _{7n+i}^{7n+j}dx_{2n+i}+X^{2n+i}\frac{\partial ^{2}L}{\partial
x_{7n+j}\partial x_{n+i}}dx_{7n+j}$

$+X^{7n+i}\frac{\partial ^{2}L}{\partial x_{7n+j}\partial x_{7n+i}}\delta
_{7n+i}^{7n+j}dx_{3n+i}-X^{3n+i}\frac{\partial ^{2}L}{\partial
x_{7n+j}\partial x_{7n+i}}dx_{7n+j}+X^{7n+i}\frac{\partial ^{2}L}{\partial
x_{7n+j}\partial x_{i}}\delta _{7n+i}^{7n+j}dx_{4n+i}\ $

$-X^{4n+i}\frac{\partial ^{2}L}{\partial x_{7n+j}\partial x_{i}}%
dx_{7n+j}-X^{7n+i}\frac{\partial ^{2}L}{\partial x_{7n+j}\partial x_{6n+i}}%
\delta _{7n+i}^{7n+j}dx_{5n+i}+X^{5n+i}\frac{\partial ^{2}L}{\partial
x_{7n+j}\partial x_{6n+i}}dx_{7n+j}$

$+X^{7n+i}\frac{\partial ^{2}L}{\partial x_{7n+j}\partial x_{5n+i}}\delta
_{7n+i}^{7n+j}dx_{6n+i}-X^{6n+i}\frac{\partial ^{2}L}{\partial
x_{7n+j}\partial x_{5n+i}}dx_{7n+j}-X^{7n+i}\frac{\partial ^{2}L}{\partial
x_{7n+j}\partial x_{3n+i}}\delta _{7n+i}^{7n+j}dx_{7n+i}$

$+X^{7n+i}\frac{\partial ^{2}L}{\partial x_{7n+j}\partial x_{3n+i}}dx_{7n+j}$

Since the closed Clifford K\"{a}hler form $\Phi _{L}^{J_{4}}$ on $(M,V)$ is
the symplectic structure, it holds

$E_{L}^{J_{4}}=V_{J_{4}}(L)-L=X^{i}\frac{\partial L}{\partial x_{4n+i}}%
-X^{n+i}\frac{\partial L}{\partial x_{2n+i}}+X^{2n+i}\frac{\partial L}{%
\partial x_{n+i}}-X^{3n+i}\frac{\partial L}{\partial x_{7n+i}}$

$\ \ \ \ \ \ \ \ \ \ \ \ \ \ \ \ \ \ \ \ \ \ \ \ \ \ \ \ -X^{4n+i}\frac{%
\partial L}{\partial x_{i}}+X^{5n+i}\frac{\partial L}{\partial x_{6n+i}}%
-X^{6n+i}\frac{\partial L}{\partial x_{5n+i}}+X^{7n+i}\frac{\partial L}{%
\partial x_{3n+i}}-L$

and thus

$dE_{L}^{J_{4}}=X^{i}\frac{\partial ^{2}L}{\partial x_{j}\partial x_{4n+i}}%
dx_{j}-X^{n+i}\frac{\partial ^{2}L}{\partial x_{j}\partial x_{2n+i}}%
dx_{j}+X^{2n+i}\frac{\partial ^{2}L}{\partial x_{j}\partial x_{n+i}}dx_{j}$

$-X^{3n+i}\frac{\partial ^{2}L}{\partial x_{j}\partial x_{7n+i}}%
dx_{j}-X^{4n+i}\frac{\partial ^{2}L}{\partial x_{j}\partial x_{i}}%
dx_{j}+X^{5n+i}\frac{\partial ^{2}L}{\partial x_{j}\partial x_{6n+i}}dx_{j}$

$-X^{6n+i}\frac{\partial ^{2}L}{\partial x_{j}\partial x_{5n+i}}%
dx_{j}+X^{7n+i}\frac{\partial ^{2}L}{\partial x_{j}\partial x_{3n+i}}%
dx_{j}+X^{i}\frac{\partial ^{2}L}{\partial x_{n+j}\partial x_{4n+i}}dx_{n+j}$

$-X^{n+i}\frac{\partial ^{2}L}{\partial x_{n+j}\partial x_{2n+i}}%
dx_{n+j}+X^{2n+i}\frac{\partial ^{2}L}{\partial x_{n+j}\partial x_{n+i}}%
dx_{n+j}-X^{3n+i}\frac{\partial ^{2}L}{\partial x_{n+j}\partial x_{7n+i}}%
dx_{n+j}$

$-X^{4n+i}\frac{\partial ^{2}L}{\partial x_{n+j}\partial x_{i}}%
dx_{n+j}+X^{5n+i}\frac{\partial ^{2}L}{\partial x_{n+j}\partial x_{6n+i}}%
dx_{n+j}-X^{6n+i}\frac{\partial ^{2}L}{\partial x_{n+j}\partial x_{5n+i}}%
dx_{n+j}$

$+X^{7n+i}\frac{\partial ^{2}L}{\partial x_{n+j}\partial x_{3n+i}}%
dx_{n+j}+X^{i}\frac{\partial ^{2}L}{\partial x_{2n+j}\partial x_{4n+i}}%
dx_{2n+j}-X^{n+i}\frac{\partial ^{2}L}{\partial x_{2n+j}\partial x_{2n+i}}%
dx_{2n+j}$

$+X^{2n+i}\frac{\partial ^{2}L}{\partial x_{2n+j}\partial x_{n+i}}%
dx_{2n+j}-X^{3n+i}\frac{\partial ^{2}L}{\partial x_{2n+j}\partial x_{7n+i}}%
dx_{2n+j}-X^{4n+i}\frac{\partial ^{2}L}{\partial x_{2n+j}\partial x_{i}}%
dx_{2n+j}$

$+X^{5n+i}\frac{\partial ^{2}L}{\partial x_{2n+j}\partial x_{6n+i}}%
dx_{2n+j}-X^{6n+i}\frac{\partial ^{2}L}{\partial x_{2n+j}\partial x_{5n+i}}%
dx_{2n+j}+X^{7n+i}\frac{\partial ^{2}L}{\partial x_{2n+j}\partial x_{3n+i}}%
dx_{2n+j}$

$+X^{i}\frac{\partial ^{2}L}{\partial x_{3n+j}\partial x_{4n+i}}%
dx_{3n+j}-X^{n+i}\frac{\partial ^{2}L}{\partial x_{3n+j}\partial x_{2n+i}}%
dx_{3n+j}+X^{2n+i}\frac{\partial ^{2}L}{\partial x_{3n+j}\partial x_{n+i}}%
dx_{3n+j}$

$-X^{3n+i}\frac{\partial ^{2}L}{\partial x_{3n+j}\partial x_{7n+i}}%
dx_{3n+j}-X^{4n+i}\frac{\partial ^{2}L}{\partial x_{3n+j}\partial x_{i}}%
dx_{3n+j}+X^{5n+i}\frac{\partial ^{2}L}{\partial x_{3n+j}\partial x_{6n+i}}%
dx_{3n+j}$

$-X^{6n+i}\frac{\partial ^{2}L}{\partial x_{3n+j}\partial x_{5n+i}}%
dx_{3n+j}+X^{7n+i}\frac{\partial ^{2}L}{\partial x_{3n+j}\partial x_{3n+i}}%
dx_{3n+j}+X^{i}\frac{\partial ^{2}L}{\partial x_{4n+j}\partial x_{4n+i}}%
dx_{4n+j}$

$-X^{n+i}\frac{\partial ^{2}L}{\partial x_{4n+j}\partial x_{2n+i}}%
dx_{4n+j}+X^{2n+i}\frac{\partial ^{2}L}{\partial x_{4n+j}\partial x_{n+i}}%
dx_{4n+j}-X^{3n+i}\frac{\partial ^{2}L}{\partial x_{4n+j}\partial x_{7n+i}}%
dx_{4n+j}$

$-X^{4n+i}\frac{\partial ^{2}L}{\partial x_{4n+j}\partial x_{i}}%
dx_{4n+j}+X^{5n+i}\frac{\partial ^{2}L}{\partial x_{4n+j}\partial x_{6n+i}}%
dx_{4n+j}-X^{6n+i}\frac{\partial ^{2}L}{\partial x_{4n+j}\partial x_{5n+i}}%
dx_{4n+j}$

$+X^{7n+i}\frac{\partial ^{2}L}{\partial x_{4n+j}\partial x_{3n+i}}%
dx_{4n+j}+X^{i}\frac{\partial ^{2}L}{\partial x_{5n+j}\partial x_{4n+i}}%
dx_{5n+j}-X^{n+i}\frac{\partial ^{2}L}{\partial x_{5n+j}\partial x_{2n+i}}%
dx_{5n+j}$

$+X^{2n+i}\frac{\partial ^{2}L}{\partial x_{5n+j}\partial x_{n+i}}%
dx_{5n+j}-X^{3n+i}\frac{\partial ^{2}L}{\partial x_{5n+j}\partial x_{7n+i}}%
dx_{5n+j}-X^{4n+i}\frac{\partial ^{2}L}{\partial x_{5n+j}\partial x_{i}}%
dx_{5n+j}$

$+X^{5n+i}\frac{\partial ^{2}L}{\partial x_{5n+j}\partial x_{6n+i}}%
dx_{5n+j}-X^{6n+i}\frac{\partial ^{2}L}{\partial x_{5n+j}\partial x_{5n+i}}%
dx_{5n+j}+X^{7n+i}\frac{\partial ^{2}L}{\partial x_{5n+j}\partial x_{3n+i}}%
dx_{5n+j}$

$+X^{i}\frac{\partial ^{2}L}{\partial x_{6n+j}\partial x_{4n+i}}%
dx_{6n+j}-X^{n+i}\frac{\partial ^{2}L}{\partial x_{6n+j}\partial x_{2n+i}}%
dx_{6n+j}+X^{2n+i}\frac{\partial ^{2}L}{\partial x_{6n+j}\partial x_{n+i}}%
dx_{6n+j}$

$-X^{3n+i}\frac{\partial ^{2}L}{\partial x_{6n+j}\partial x_{7n+i}}%
dx_{6n+j}-X^{4n+i}\frac{\partial ^{2}L}{\partial x_{6n+j}\partial x_{i}}%
dx_{6n+j}+X^{5n+i}\frac{\partial ^{2}L}{\partial x_{6n+j}\partial x_{6n+i}}%
dx_{6n+j}$

$-X^{6n+i}\frac{\partial ^{2}L}{\partial x_{6n+j}\partial x_{5n+i}}%
dx_{6n+j}+X^{7n+i}\frac{\partial ^{2}L}{\partial x_{6n+j}\partial x_{3n+i}}%
dx_{6n+j}+X^{i}\frac{\partial ^{2}L}{\partial x_{7n+j}\partial x_{4n+i}}%
dx_{7n+j}$

$-X^{n+i}\frac{\partial ^{2}L}{\partial x_{7n+j}\partial x_{2n+i}}%
dx_{7n+j}+X^{2n+i}\frac{\partial ^{2}L}{\partial x_{7n+j}\partial x_{n+i}}%
dx_{7n+j}-X^{3n+i}\frac{\partial ^{2}L}{\partial x_{7n+j}\partial x_{7n+i}}%
dx_{7n+j}$

$-X^{4n+i}\frac{\partial ^{2}L}{\partial x_{7n+j}\partial x_{i}}%
dx_{7n+j}+X^{5n+i}\frac{\partial ^{2}L}{\partial x_{7n+j}\partial x_{6n+i}}%
dx_{7n+j}-X^{6n+i}\frac{\partial ^{2}L}{\partial x_{7n+j}\partial x_{5n+i}}%
dx_{7n+j}$

$+X^{7n+i}\frac{\partial ^{2}L}{\partial x_{7n+j}\partial x_{3n+i}}dx_{7n+j}-%
\frac{\partial L}{\partial x_{j}}dx_{j}-\frac{\partial L}{\partial x_{n+j}}%
dx_{n+j}-\frac{\partial L}{\partial x_{2n+j}}dx_{2n+j}$

$-\frac{\partial L}{\partial x_{3n+j}}dx_{3n+j}-\frac{\partial L}{\partial
x_{4n+j}}dx_{4n+j}-\frac{\partial L}{\partial x_{5n+j}}dx_{5n+j}-\frac{%
\partial L}{\partial x_{6n+j}}dx_{6n+j}-\frac{\partial L}{\partial x_{7n+j}}%
dx_{7n+j}$

By means of \textbf{Eq.} (\ref{1.1}), we calculate the following expressions

$-X^{i}\frac{\partial ^{2}L}{\partial x_{j}\partial x_{4n+i}}\delta
_{i}^{j}dx_{i}+X^{i}\frac{\partial ^{2}L}{\partial x_{j}\partial x_{2n+i}}%
\delta _{i}^{j}dx_{n+i}-X^{i}\frac{\partial ^{2}L}{\partial x_{j}\partial
x_{n+i}}\delta _{i}^{j}dx_{2n+i}$

$+X^{i}\frac{\partial ^{2}L}{\partial x_{j}\partial x_{7n+i}}\delta
_{i}^{j}dx_{3n+i}+X^{i}\frac{\partial ^{2}L}{\partial x_{j}\partial x_{i}}%
\delta _{i}^{j}dx_{4n+i}-X^{i}\frac{\partial ^{2}L}{\partial x_{j}\partial
x_{6n+i}}\delta _{i}^{j}dx_{5n+i}$

$+X^{i}\frac{\partial ^{2}L}{\partial x_{j}\partial x_{5n+i}}\delta
_{i}^{j}dx_{6n+i}-X^{i}\frac{\partial ^{2}L}{\partial x_{j}\partial x_{3n+i}}%
\delta _{i}^{j}dx_{7n+i}-X^{n+i}\frac{\partial ^{2}L}{\partial
x_{n+j}\partial x_{4n+i}}\delta _{n+i}^{n+j}dx_{i}$

$+X^{n+i}\frac{\partial ^{2}L}{\partial x_{n+j}\partial x_{2n+i}}\delta
_{n+i}^{n+j}dx_{n+i}-X^{n+i}\frac{\partial ^{2}L}{\partial x_{n+j}\partial
x_{n+i}}\delta _{n+i}^{n+j}dx_{2n+i}\ +X^{n+i}\frac{\partial ^{2}L}{\partial
x_{n+j}\partial x_{7n+i}}\delta _{n+i}^{n+j}dx_{3n+i}$

$+X^{n+i}\frac{\partial ^{2}L}{\partial x_{n+j}\partial x_{i}}\delta
_{n+i}^{n+j}dx_{4n+i}-X^{n+i}\frac{\partial ^{2}L}{\partial x_{n+j}\partial
x_{6n+i}}\delta _{n+i}^{n+j}dx_{5n+i}+X^{n+i}\frac{\partial ^{2}L}{\partial
x_{n+j}\partial x_{5n+i}}\delta _{n+i}^{n+j}dx_{6n+i}$

$-X^{n+i}\frac{\partial ^{2}L}{\partial x_{n+j}\partial x_{7n+i}}\delta
_{n+i}^{n+j}dx_{7n+i}-\ X^{2n+i}\frac{\partial ^{2}L}{\partial
x_{2n+j}\partial x_{4n+i}}\delta _{2n+i}^{2n+j}dx_{i}+X^{2n+i}\frac{\partial
^{2}L}{\partial x_{2n+j}\partial x_{2n+i}}\delta _{2n+i}^{2n+j}dx_{n+i}$

$-X^{2n+i}\frac{\partial ^{2}L}{\partial x_{2n+j}\partial x_{n+i}}\delta
_{2n+i}^{2n+j}dx_{2n+i}+X^{2n+i}\frac{\partial ^{2}L}{\partial
x_{2n+j}\partial x_{7n+i}}\delta _{2n+i}^{2n+j}dx_{3n+i}+X^{2n+i}\frac{%
\partial ^{2}L}{\partial x_{2n+j}\partial x_{i}}\delta
_{2n+i}^{2n+j}dx_{4n+i}$

$-X^{2n+i}\frac{\partial ^{2}L}{\partial x_{2n+j}\partial x_{6n+i}}\delta
_{2n+i}^{2n+j}dx_{5n+i}+X^{2n+i}\frac{\partial ^{2}L}{\partial
x_{2n+j}\partial x_{5n+i}}\delta _{2n+i}^{2n+j}dx_{6n+i}-X^{2n+i}\frac{%
\partial ^{2}L}{\partial x_{2n+j}\partial x_{7n+i}}\delta
_{2n+i}^{2n+j}dx_{7n+i}$

$-X^{3n+i}\frac{\partial ^{2}L}{\partial x_{3n+j}\partial x_{4n+i}}\delta
_{3n+i}^{3n+j}dx_{i}+X^{3n+i}\frac{\partial ^{2}L}{\partial x_{3n+j}\partial
x_{2n+i}}\delta _{3n+i}^{3n+j}dx_{n+i}-X^{3n+i}\frac{\partial ^{2}L}{%
\partial x_{3n+j}\partial x_{n+i}}\delta _{3n+i}^{3n+j}dx_{2n+i}$

$+X^{3n+i}\frac{\partial ^{2}L}{\partial x_{3n+j}\partial x_{7n+i}}\delta
_{3n+i}^{3n+j}dx_{3n+i}+X^{3n+i}\frac{\partial ^{2}L}{\partial
x_{3n+j}\partial x_{i}}\delta _{3n+i}^{3n+j}dx_{4n+i}-X^{3n+i}\frac{\partial
^{2}L}{\partial x_{3n+j}\partial x_{6n+i}}\delta _{3n+i}^{3n+j}dx_{5n+i}$

$+X^{3n+i}\frac{\partial ^{2}L}{\partial x_{3n+j}\partial x_{5n+i}}\delta
_{3n+i}^{3n+j}dx_{6n+i}-X^{3n+i}\frac{\partial ^{2}L}{\partial
x_{3n+j}\partial x_{7n+i}}\delta _{3n+i}^{3n+j}dx_{7n+i}-X^{4n+i}\frac{%
\partial L}{\partial x_{4n+j}\partial x_{4n+i}}\delta _{4n+i}^{4n+j}dx_{i}\ $

$+X^{4n+i}\frac{\partial L}{\partial x_{4n+j}\partial x_{2n+i}}\delta
_{4n+i}^{4n+j}dx_{n+i}-X^{4n+i}\frac{\partial L}{\partial x_{4n+j}\partial
x_{n+i}}\delta _{4n+i}^{4n+j}dx_{2n+i}+X^{4n+i}\frac{\partial L}{\partial
x_{4n+j}\partial x_{7n+i}}\delta _{4n+i}^{4n+j}dx_{3n+i}$

$+X^{4n+i}\frac{\partial ^{2}L}{\partial x_{4n+j}\partial x_{i}}\delta
_{4n+i}^{4n+j}dx_{4n+i}-X^{4n+i}\frac{\partial ^{2}L}{\partial
x_{4n+j}\partial x_{6n+i}}\delta _{4n+i}^{4n+j}dx_{5n+i}+X^{4n+i}\frac{%
\partial ^{2}L}{\partial x_{4n+j}\partial x_{5n+i}}\delta
_{4n+i}^{4n+j}dx_{6n+i}$

$-X^{4n+i}\frac{\partial ^{2}L}{\partial x_{4n+j}\partial x_{3n+i}}\delta
_{4n+i}^{4n+j}dx_{7n+i}-X^{5n+i}\frac{\partial ^{2}L}{\partial
x_{5n+j}\partial x_{4n+i}}\delta _{5n+i}^{5n+j}dx_{i}+X^{5n+i}\frac{\partial
^{2}L}{\partial x_{5n+j}\partial x_{2n+i}}\delta _{5n+i}^{5n+j}dx_{n+i}$

$-X^{5n+i}\frac{\partial ^{2}L}{\partial x_{5n+j}\partial x_{n+i}}\delta
_{5n+i}^{5n+j}dx_{2n+i}+X^{5n+i}\frac{\partial ^{2}L}{\partial
x_{5n+j}\partial x_{7n+i}}\delta _{5n+i}^{5n+j}dx_{3n+i}+X^{5n+i}\frac{%
\partial ^{2}L}{\partial x_{5n+j}\partial x_{i}}\delta
_{5n+i}^{5n+j}dx_{4n+i}$

$-X^{5n+i}\frac{\partial ^{2}L}{\partial x_{5n+j}\partial x_{6n+i}}\delta
_{5n+i}^{5n+j}dx_{5n+i}+X^{5n+i}\frac{\partial ^{2}L}{\partial
x_{5n+j}\partial x_{5n+i}}\delta _{5n+i}^{5n+j}dx_{6n+i}-X^{5n+i}\frac{%
\partial ^{2}L}{\partial x_{5n+j}\partial x_{3n+i}}\delta
_{5n+i}^{5n+j}dx_{7n+i}$

$-X^{6n+i}\frac{\partial ^{2}L}{\partial x_{6n+j}\partial x_{4n+i}}\delta
_{6n+i}^{6n+j}dx_{i}+X^{6n+i}\frac{\partial ^{2}L}{\partial x_{6n+j}\partial
x_{2n+i}}\delta _{6n+i}^{6n+j}dx_{n+i}-X^{6n+i}\frac{\partial ^{2}L}{%
\partial x_{6n+j}\partial x_{n+i}}\delta _{6n+i}^{6n+j}dx_{2n+i}$

$+X^{6n+i}\frac{\partial ^{2}L}{\partial x_{6n+j}\partial x_{7n+i}}\delta
_{6n+i}^{6n+j}dx_{3n+i}+X^{6n+i}\frac{\partial ^{2}L}{\partial
x_{6n+j}\partial x_{i}}\delta _{6n+i}^{6n+j}dx_{4n+i}-X^{6n+i}\frac{\partial
^{2}L}{\partial x_{6n+j}\partial x_{6n+i}}\delta _{6n+i}^{6n+j}dx_{5n+i}$

$+X^{6n+i}\frac{\partial ^{2}L}{\partial x_{6n+j}\partial x_{5n+i}}\delta
_{6n+i}^{6n+j}dx_{6n+i}-X^{6n+i}\frac{\partial ^{2}L}{\partial
x_{6n+j}\partial x_{7n+i}}\delta _{6n+i}^{6n+j}dx_{7n+i}-X^{7n+i}\frac{%
\partial ^{2}L}{\partial x_{7n+j}\partial x_{4n+i}}\delta
_{7n+i}^{7n+j}dx_{i}$

$+X^{7n+i}\frac{\partial ^{2}L}{\partial x_{7n+j}\partial x_{2n+i}}\delta
_{7n+i}^{7n+j}dx_{n+i}-X^{7n+i}\frac{\partial ^{2}L}{\partial
x_{7n+j}\partial x_{n+i}}\delta _{7n+i}^{7n+j}dx_{2n+i}+X^{7n+i}\frac{%
\partial ^{2}L}{\partial x_{7n+j}\partial x_{7n+i}}\delta
_{7n+i}^{7n+j}dx_{3n+i}$

$+X^{7n+i}\frac{\partial ^{2}L}{\partial x_{7n+j}\partial x_{i}}\delta
_{7n+i}^{7n+j}dx_{4n+i}-X^{7n+i}\frac{\partial ^{2}L}{\partial
x_{7n+j}\partial x_{6n+i}}\delta _{7n+i}^{7n+j}dx_{5n+i}+X^{7n+i}\frac{%
\partial ^{2}L}{\partial x_{7n+j}\partial x_{5n+i}}\delta
_{7n+i}^{7n+j}dx_{6n+i}$

$-X^{7n+i}\frac{\partial ^{2}L}{\partial x_{7n+j}\partial x_{3n+i}}\delta
_{7n+i}^{7n+j}dx_{7n+i}+\frac{\partial L}{\partial x_{j}}dx_{j}+\frac{%
\partial L}{\partial x_{n+j}}dx_{n+j}+\frac{\partial L}{\partial x_{2n+j}}%
dx_{2n+j}+\frac{\partial L}{\partial x_{3n+j}}dx_{3n+j}$

$+\frac{\partial L}{\partial x_{4n+j}}dx_{4n+j}+\frac{\partial L}{\partial
x_{5n+j}}dx_{5n+j}+\frac{\partial L}{\partial x_{6n+j}}dx_{6n+j}+\frac{%
\partial L}{\partial x_{7n+j}}dx_{7n+j}=0.$

If a curve determined by $\alpha :\mathbf{R}\rightarrow M$ is taken to be an
integral curve of $\xi ,$ then we found equation as follows:

$-X^{i}\frac{\partial ^{2}L}{\partial x_{j}\partial x_{4n+i}}dx_{j}-X^{n+i}%
\frac{\partial ^{2}L}{\partial x_{n+j}\partial x_{4n+i}}dx_{j}-\ X^{2n+i}%
\frac{\partial ^{2}L}{\partial x_{2n+j}\partial x_{4n+i}}dx_{j}$

$-X^{3n+i}\frac{\partial ^{2}L}{\partial x_{3n+j}\partial x_{4n+i}}%
dx_{j}-X^{4n+i}\frac{\partial L}{\partial x_{4n+j}\partial x_{4n+i}}%
dx_{j}-X^{5n+i}\frac{\partial ^{2}L}{\partial x_{5n+j}\partial x_{4n+i}}%
dx_{j}$

$-X^{6n+i}\frac{\partial ^{2}L}{\partial x_{6n+j}\partial x_{4n+i}}%
dx_{j}-X^{7n+i}\frac{\partial ^{2}L}{\partial x_{7n+j}\partial x_{4n+i}}%
dx_{j}+X^{i}\frac{\partial ^{2}L}{\partial x_{j}\partial x_{2n+i}}dx_{n+j}$

$+X^{n+i}\frac{\partial ^{2}L}{\partial x_{n+j}\partial x_{2n+i}}%
dx_{n+j}+X^{2n+i}\frac{\partial ^{2}L}{\partial x_{2n+j}\partial x_{2n+i}}%
dx_{n+j}+X^{3n+i}\frac{\partial ^{2}L}{\partial x_{3n+j}\partial x_{2n+i}}%
dx_{n+j}$

$+X^{4n+i}\frac{\partial L}{\partial x_{4n+j}\partial x_{2n+i}}%
dx_{n+j}+X^{5n+i}\frac{\partial ^{2}L}{\partial x_{5n+j}\partial x_{2n+i}}%
dx_{n+j}+X^{6n+i}\frac{\partial ^{2}L}{\partial x_{6n+j}\partial x_{2n+i}}%
dx_{n+j}$

$+X^{7n+i}\frac{\partial ^{2}L}{\partial x_{7n+j}\partial x_{2n+i}}%
dx_{n+j}-X^{i}\frac{\partial ^{2}L}{\partial x_{j}\partial x_{n+i}}%
dx_{2n+j}-X^{n+i}\frac{\partial ^{2}L}{\partial x_{n+j}\partial x_{n+i}}%
dx_{2n+j}\ $

$-X^{2n+i}\frac{\partial ^{2}L}{\partial x_{2n+j}\partial x_{n+i}}%
dx_{2n+j}-X^{3n+i}\frac{\partial ^{2}L}{\partial x_{3n+j}\partial x_{n+i}}%
dx_{2n+j}-X^{4n+i}\frac{\partial L}{\partial x_{4n+j}\partial x_{n+i}}%
dx_{2n+j}$

$-X^{5n+i}\frac{\partial ^{2}L}{\partial x_{5n+j}\partial x_{n+i}}%
dx_{2n+j}-X^{6n+i}\frac{\partial ^{2}L}{\partial x_{6n+j}\partial x_{n+i}}%
dx_{2n+j}-X^{7n+i}\frac{\partial ^{2}L}{\partial x_{7n+j}\partial x_{n+i}}%
dx_{2n+j}$

$+X^{i}\frac{\partial ^{2}L}{\partial x_{j}\partial x_{7n+i}}%
dx_{3n+j}+X^{n+i}\frac{\partial ^{2}L}{\partial x_{n+j}\partial x_{7n+i}}%
dx_{3n+j}+X^{2n+i}\frac{\partial ^{2}L}{\partial x_{2n+j}\partial x_{7n+i}}%
dx_{3n+j}$

$+X^{3n+i}\frac{\partial ^{2}L}{\partial x_{3n+j}\partial x_{7n+i}}%
dx_{3n+j}+X^{4n+i}\frac{\partial L}{\partial x_{4n+j}\partial x_{7n+i}}%
dx_{3n+j}+X^{5n+i}\frac{\partial ^{2}L}{\partial x_{5n+j}\partial x_{7n+i}}%
dx_{3n+j}$

$+X^{6n+i}\frac{\partial ^{2}L}{\partial x_{6n+j}\partial x_{7n+i}}%
dx_{3n+j}+X^{7n+i}\frac{\partial ^{2}L}{\partial x_{7n+j}\partial x_{7n+i}}%
dx_{3n+j}+X^{i}\frac{\partial ^{2}L}{\partial x_{j}\partial x_{i}}dx_{4n+j}$

$+X^{n+i}\frac{\partial ^{2}L}{\partial x_{n+j}\partial x_{i}}%
dx_{4n+j}+X^{2n+i}\frac{\partial ^{2}L}{\partial x_{2n+j}\partial x_{i}}%
dx_{4n+j}+X^{3n+i}\frac{\partial ^{2}L}{\partial x_{3n+j}\partial x_{i}}%
dx_{4n+j}$

$+X^{4n+i}\frac{\partial ^{2}L}{\partial x_{4n+j}\partial x_{i}}%
dx_{4n+j}+X^{5n+i}\frac{\partial ^{2}L}{\partial x_{5n+j}\partial x_{i}}%
dx_{4n+j}+X^{3n+i}\frac{\partial ^{2}L}{\partial x_{3n+j}\partial x_{i}}%
dx_{4n+j}$

$+X^{7n+i}\frac{\partial ^{2}L}{\partial x_{7n+j}\partial x_{i}}%
dx_{4n+j}-X^{i}\frac{\partial ^{2}L}{\partial x_{j}\partial x_{6n+i}}%
dx_{5n+j}-X^{n+i}\frac{\partial ^{2}L}{\partial x_{n+j}\partial x_{6n+i}}%
dx_{5n+j}$

$-X^{2n+i}\frac{\partial ^{2}L}{\partial x_{2n+j}\partial x_{6n+i}}%
dx_{5n+j}-X^{3n+i}\frac{\partial ^{2}L}{\partial x_{3n+j}\partial x_{6n+i}}%
dx_{5n+j}-X^{4n+i}\frac{\partial ^{2}L}{\partial x_{4n+j}\partial x_{6n+i}}%
dx_{5n+j}$

$-X^{5n+i}\frac{\partial ^{2}L}{\partial x_{5n+j}\partial x_{6n+i}}%
dx_{5n+j}-X^{6n+i}\frac{\partial ^{2}L}{\partial x_{6n+j}\partial x_{6n+i}}%
dx_{5n+j}-X^{7n+i}\frac{\partial ^{2}L}{\partial x_{7n+j}\partial x_{6n+i}}%
dx_{5n+j}$

$+X^{i}\frac{\partial ^{2}L}{\partial x_{j}\partial x_{5n+i}}%
dx_{6n+j}+X^{n+i}\frac{\partial ^{2}L}{\partial x_{n+j}\partial x_{5n+i}}%
dx_{6n+j}+X^{2n+i}\frac{\partial ^{2}L}{\partial x_{2n+j}\partial x_{5n+i}}%
dx_{6n+j}$

$+X^{3n+i}\frac{\partial ^{2}L}{\partial x_{3n+j}\partial x_{5n+i}}%
dx_{6n+j}+X^{4n+i}\frac{\partial ^{2}L}{\partial x_{4n+j}\partial x_{5n+i}}%
dx_{6n+j}+X^{5n+i}\frac{\partial ^{2}L}{\partial x_{5n+j}\partial x_{5n+i}}%
dx_{6n+j}$

$+X^{6n+i}\frac{\partial ^{2}L}{\partial x_{6n+j}\partial x_{5n+i}}%
dx_{6n+j}+X^{7n+i}\frac{\partial ^{2}L}{\partial x_{7n+j}\partial x_{5n+i}}%
dx_{6n+j}-X^{i}\frac{\partial ^{2}L}{\partial x_{j}\partial x_{3n+i}}%
dx_{7n+j}$

$-X^{n+i}\frac{\partial ^{2}L}{\partial x_{n+j}\partial x_{7n+i}}%
dx_{7n+j}-X^{2n+i}\frac{\partial ^{2}L}{\partial x_{2n+j}\partial x_{7n+i}}%
dx_{7n+j}-X^{3n+i}\frac{\partial ^{2}L}{\partial x_{3n+j}\partial x_{7n+i}}%
dx_{7n+j}$

$-X^{4n+i}\frac{\partial ^{2}L}{\partial x_{4n+j}\partial x_{3n+i}}%
dx_{7n+j}-X^{5n+i}\frac{\partial ^{2}L}{\partial x_{5n+j}\partial x_{3n+i}}%
dx_{7n+j}-X^{6n+i}\frac{\partial ^{2}L}{\partial x_{6n+j}\partial x_{7n+i}}%
dx_{7n+j}$

$-X^{7n+i}\frac{\partial ^{2}L}{\partial x_{7n+j}\partial x_{3n+i}}dx_{7n+j}+%
\frac{\partial L}{\partial x_{j}}dx_{j}+\frac{\partial L}{\partial x_{n+j}}%
dx_{n+j}+\frac{\partial L}{\partial x_{2n+j}}dx_{2n+j}+\frac{\partial L}{%
\partial x_{3n+j}}dx_{3n+j}$

$+\frac{\partial L}{\partial x_{4n+j}}dx_{4n+j}+\frac{\partial L}{\partial
x_{5n+j}}dx_{5n+j}+\frac{\partial L}{\partial x_{6n+j}}dx_{6n+j}+\frac{%
\partial L}{\partial x_{7n+j}}dx_{7n+j}=0$

\textbf{\ }or

$-[X^{i}\frac{\partial ^{2}L}{\partial x_{j}\partial x_{4n+i}}+X^{n+i}\frac{%
\partial ^{2}L}{\partial x_{n+j}\partial x_{4n+i}}+\ X^{2n+i}\frac{\partial
^{2}L}{\partial x_{2n+j}\partial x_{4n+i}}+X^{3n+i}\frac{\partial ^{2}L}{%
\partial x_{3n+j}\partial x_{4n+i}}+X^{4n+i}\frac{\partial L}{\partial
x_{4n+j}\partial x_{4n+i}}$

$+X^{5n+i}\frac{\partial ^{2}L}{\partial x_{5n+j}\partial x_{4n+i}}+X^{6n+i}%
\frac{\partial ^{2}L}{\partial x_{6n+j}\partial x_{4n+i}}+X^{7n+i}\frac{%
\partial ^{2}L}{\partial x_{7n+j}\partial x_{4n+i}}]dx_{j}+\frac{\partial L}{%
\partial x_{j}}dx_{j}\ $

$+[X^{i}\frac{\partial ^{2}L}{\partial x_{j}\partial x_{2n+i}}+X^{n+i}\frac{%
\partial ^{2}L}{\partial x_{n+j}\partial x_{2n+i}}+X^{2n+i}\frac{\partial
^{2}L}{\partial x_{2n+j}\partial x_{2n+i}}+X^{3n+i}\frac{\partial ^{2}L}{%
\partial x_{3n+j}\partial x_{2n+i}}+X^{4n+i}\frac{\partial L}{\partial
x_{4n+j}\partial x_{2n+i}}$

$+X^{5n+i}\frac{\partial ^{2}L}{\partial x_{5n+j}\partial x_{2n+i}}+X^{6n+i}%
\frac{\partial ^{2}L}{\partial x_{6n+j}\partial x_{2n+i}}+X^{7n+i}\frac{%
\partial ^{2}L}{\partial x_{7n+j}\partial x_{2n+i}}]dx_{n+j}+\frac{\partial L%
}{\partial x_{n+j}}dx_{n+j}$

$-[X^{i}\frac{\partial ^{2}L}{\partial x_{j}\partial x_{n+i}}+X^{n+i}\frac{%
\partial ^{2}L}{\partial x_{n+j}\partial x_{n+i}}+X^{2n+i}\frac{\partial
^{2}L}{\partial x_{2n+j}\partial x_{n+i}}+X^{3n+i}\frac{\partial ^{2}L}{%
\partial x_{3n+j}\partial x_{n+i}}+X^{4n+i}\frac{\partial L}{\partial
x_{4n+j}\partial x_{n+i}}$

$+X^{5n+i}\frac{\partial ^{2}L}{\partial x_{5n+j}\partial x_{n+i}}+X^{6n+i}%
\frac{\partial ^{2}L}{\partial x_{6n+j}\partial x_{n+i}}+X^{7n+i}\frac{%
\partial ^{2}L}{\partial x_{7n+j}\partial x_{n+i}}]dx_{2n+j}+\frac{\partial L%
}{\partial x_{2n+j}}dx_{2n+j}$

$+[X^{i}\frac{\partial ^{2}L}{\partial x_{j}\partial x_{7n+i}}+X^{n+i}\frac{%
\partial ^{2}L}{\partial x_{n+j}\partial x_{7n+i}}+X^{2n+i}\frac{\partial
^{2}L}{\partial x_{2n+j}\partial x_{7n+i}}+X^{3n+i}\frac{\partial ^{2}L}{%
\partial x_{3n+j}\partial x_{7n+i}}+X^{4n+i}\frac{\partial L}{\partial
x_{4n+j}\partial x_{7n+i}}$

$+X^{5n+i}\frac{\partial ^{2}L}{\partial x_{5n+j}\partial x_{7n+i}}+X^{6n+i}%
\frac{\partial ^{2}L}{\partial x_{6n+j}\partial x_{7n+i}}+X^{7n+i}\frac{%
\partial ^{2}L}{\partial x_{7n+j}\partial x_{7n+i}}]dx_{3n+j}+\frac{\partial
L}{\partial x_{3n+j}}dx_{3n+j}$

$+[X^{i}\frac{\partial ^{2}L}{\partial x_{j}\partial x_{i}}+X^{n+i}\frac{%
\partial ^{2}L}{\partial x_{n+j}\partial x_{i}}+X^{2n+i}\frac{\partial ^{2}L%
}{\partial x_{2n+j}\partial x_{i}}+X^{3n+i}\frac{\partial ^{2}L}{\partial
x_{3n+j}\partial x_{i}}+X^{4n+i}\frac{\partial ^{2}L}{\partial
x_{4n+j}\partial x_{i}}$

$+X^{5n+i}\frac{\partial ^{2}L}{\partial x_{5n+j}\partial x_{i}}+X^{3n+i}%
\frac{\partial ^{2}L}{\partial x_{3n+j}\partial x_{i}}+X^{7n+i}\frac{%
\partial ^{2}L}{\partial x_{7n+j}\partial x_{i}}]dx_{4n+j}+\frac{\partial L}{%
\partial x_{4n+j}}dx_{4n+j}$

$-[X^{i}\frac{\partial ^{2}L}{\partial x_{j}\partial x_{6n+i}}+X^{n+i}\frac{%
\partial ^{2}L}{\partial x_{n+j}\partial x_{6n+i}}+X^{2n+i}\frac{\partial
^{2}L}{\partial x_{2n+j}\partial x_{6n+i}}+X^{3n+i}\frac{\partial ^{2}L}{%
\partial x_{3n+j}\partial x_{6n+i}}+X^{4n+i}\frac{\partial ^{2}L}{\partial
x_{4n+j}\partial x_{6n+i}}$

$+X^{5n+i}\frac{\partial ^{2}L}{\partial x_{5n+j}\partial x_{6n+i}}+X^{6n+i}%
\frac{\partial ^{2}L}{\partial x_{6n+j}\partial x_{6n+i}}+X^{7n+i}\frac{%
\partial ^{2}L}{\partial x_{7n+j}\partial x_{6n+i}}]dx_{5n+j}+\frac{\partial
L}{\partial x_{5n+j}}dx_{5n+j}$

$+[X^{i}\frac{\partial ^{2}L}{\partial x_{j}\partial x_{5n+i}}+X^{n+i}\frac{%
\partial ^{2}L}{\partial x_{n+j}\partial x_{5n+i}}+X^{2n+i}\frac{\partial
^{2}L}{\partial x_{2n+j}\partial x_{5n+i}}+X^{3n+i}\frac{\partial ^{2}L}{%
\partial x_{3n+j}\partial x_{5n+i}}+X^{4n+i}\frac{\partial ^{2}L}{\partial
x_{4n+j}\partial x_{5n+i}}$

$+X^{5n+i}\frac{\partial ^{2}L}{\partial x_{5n+j}\partial x_{5n+i}}+X^{6n+i}%
\frac{\partial ^{2}L}{\partial x_{6n+j}\partial x_{5n+i}}+X^{7n+i}\frac{%
\partial ^{2}L}{\partial x_{7n+j}\partial x_{5n+i}}]dx_{6n+j}+\frac{\partial
L}{\partial x_{6n+j}}dx_{6n+j}$

$-[X^{i}\frac{\partial ^{2}L}{\partial x_{j}\partial x_{3n+i}}+X^{n+i}\frac{%
\partial ^{2}L}{\partial x_{n+j}\partial x_{7n+i}}+X^{2n+i}\frac{\partial
^{2}L}{\partial x_{2n+j}\partial x_{7n+i}}+X^{3n+i}\frac{\partial ^{2}L}{%
\partial x_{3n+j}\partial x_{7n+i}}+X^{4n+i}\frac{\partial ^{2}L}{\partial
x_{4n+j}\partial x_{3n+i}}$

$+X^{5n+i}\frac{\partial ^{2}L}{\partial x_{5n+j}\partial x_{3n+i}}+X^{6n+i}%
\frac{\partial ^{2}L}{\partial x_{6n+j}\partial x_{7n+i}}+X^{7n+i}\frac{%
\partial ^{2}L}{\partial x_{7n+j}\partial x_{3n+i}}]dx_{7n+j}+\frac{\partial
L}{\partial x_{7n+j}}dx_{7n+j}=0$

Then we find the equations

\begin{equation}
\begin{array}{c}
\frac{\partial }{\partial t}\left( \frac{\partial L}{\partial x_{i}}\right) +%
\frac{\partial L}{\partial x_{4n+i}}=0,\frac{\partial }{\partial t}\left(
\frac{\partial L}{\partial x_{n+i}}\right) -\frac{\partial L}{\partial
x_{2n+i}}=0,\frac{\partial }{\partial t}\left( \frac{\partial L}{\partial
x_{2n+i}}\right) +\frac{\partial L}{\partial x_{n+i}}=0, \\
\frac{\partial }{\partial t}\left( \frac{\partial L}{\partial x_{3n+i}}%
\right) -\frac{\partial L}{\partial x_{7n+i}}=0,\frac{\partial }{\partial t}%
\left( \frac{\partial L}{\partial x_{4n+i}}\right) -\frac{\partial L}{%
\partial x_{i}}=0,\frac{\partial }{\partial t}\left( \frac{\partial L}{%
\partial x_{5n+i}}\right) +\frac{\partial L}{\partial x_{6n+i}}=0, \\
\frac{\partial }{\partial t}\left( \frac{\partial L}{\partial x_{6n+i}}%
\right) -\frac{\partial L}{\partial x_{5n+i}}=0,\frac{\partial }{\partial t}%
\left( \frac{\partial L}{\partial x_{7n+i}}\right) +\frac{\partial L}{%
\partial x_{3n+i}}=0,%
\end{array}
\label{3.6}
\end{equation}

such that the equations expressed in \textbf{Eq. }(\ref{3.6}) are named
\textit{Euler-Lagrange equations} structured on Clifford K\"{a}hler manifold
$(M,V)$ by means of $\Phi _{L}^{J_{4}}$ and in the case, the triple $(M,\Phi
_{L}^{J_{4}},\xi )$ is said to be a \textit{mechanical system }on Clifford K%
\"{a}hler manifold $(M,V)$\textit{.}

Fifth, we obtain Euler-Lagrange equations for quantum and classical
mechanics by means of $\Phi _{L}^{J_{5}}$ on Clifford K\"{a}hler manifold $%
(M,V).$

Let $J_{5}$ be another local basis component on the Clifford K\"{a}hler
manifold $(M,V).$ Let $\xi $ take as in \textbf{Eq.} (\ref{3.1}). In the
case, the vector field defined by

\begin{equation}
\begin{array}{c}
V_{J_{5}}=J_{5}(\xi )=X^{i}\frac{\partial }{\partial x_{5n+i}}-X^{n+i}\frac{%
\partial }{\partial x_{3n+i}}-X^{2n+i}\frac{\partial }{\partial x_{7n+i}}%
+X^{3n+i}\frac{\partial }{\partial x_{n+i}} \\
+X^{4n+i}\frac{\partial }{\partial x_{6n+i}}-X^{5n+i}\frac{\partial }{%
\partial x_{i}}-X^{6n+i}\frac{\partial }{\partial x_{4n+i}}+X^{7n+i}\frac{%
\partial }{\partial x_{2n+i}},%
\end{array}
\label{3.7}
\end{equation}

is \textit{Liouville vector field} on Clifford K\"{a}hler manifold $(M,V)$.
The function given by $E_{L}^{J_{5}}=V_{J_{5}}(L)-L$ is\textit{\ energy
function}. Then the operator $i_{J_{5}}$ induced by $J_{5}$ and defined by%
\begin{equation}
i_{J_{5}}\omega (X_{1},X_{2},...,X_{r})=\sum_{i=1}^{r}\omega
(X_{1},...,J_{5}X_{i},...,X_{r})  \label{3.8}
\end{equation}%
is \textit{vertical derivation, }where $\omega \in \wedge ^{r}{}M,$ $%
X_{i}\in \chi (M).$ The \textit{vertical differentiation} $d_{J_{5}}$ is
determined by%
\begin{equation}
d_{J_{5}}=[i_{J_{5}},d]=i_{J_{5}}d-di_{J_{5}}.  \label{3.9}
\end{equation}%
Taking into consideration $J_{5},$ the closed Clifford K\"{a}hler form is
the closed 2-form given by $\Phi _{L}^{J_{5}}=-dd_{J_{5}}L$ such that%
\begin{eqnarray}
d_{_{J_{5}}} &=&\frac{\partial }{\partial x_{5n+i}}dx_{i}-\frac{\partial }{%
\partial x_{3n+i}}dx_{n+i}-\frac{\partial }{\partial x_{7n+i}}dx_{2n+i}+%
\frac{\partial }{\partial x_{n+i}}dx_{3n+i}  \label{3.10} \\
&&+\frac{\partial }{\partial x_{6n+i}}dx_{4n+i}-\frac{\partial }{\partial
x_{i}}dx_{5n+i}-\frac{\partial }{\partial x_{4n+i}}dx_{6n+i}+\frac{\partial
}{\partial x_{2n+i}}dx_{7n+i}  \notag
\end{eqnarray}

and given by operator%
\begin{equation}
d_{J_{5}}:\mathcal{F}(M)\rightarrow \wedge ^{1}{}M.  \label{3.11}
\end{equation}

The closed Clifford K\"{a}hler form $\Phi _{L}^{J_{5}}$ on $M$ is the
symplectic structure. So it yields

\begin{eqnarray}
E_{L}^{J_{5}} &=&V_{J_{5}}(L)-L=X^{i}\frac{\partial L}{\partial x_{5n+i}}%
-X^{n+i}\frac{\partial L}{\partial x_{3n+i}}-X^{2n+i}\frac{\partial L}{%
\partial x_{7n+i}}+X^{3n+i}\frac{\partial L}{\partial x_{n+i}}  \label{3.12}
\\
&&+X^{4n+i}\frac{\partial L}{\partial x_{6n+i}}-X^{5n+i}\frac{\partial L}{%
\partial x_{i}}-X^{6n+i}\frac{\partial L}{\partial x_{4n+i}}+X^{7n+i}\frac{%
\partial L}{\partial x_{2n+i}}-L  \notag
\end{eqnarray}

Using \textbf{Eq.} (\ref{1.1}), using (\ref{3.1}), (\ref{3.10}) and (\ref%
{3.12}), also by means of the above fourth part we obtain the equations
\begin{equation}
\begin{array}{c}
\frac{\partial }{\partial t}\left( \frac{\partial L}{\partial x_{i}}\right) +%
\frac{\partial L}{\partial x_{5n+i}}=0,\frac{\partial }{\partial t}\left(
\frac{\partial L}{\partial x_{n+i}}\right) -\frac{\partial L}{\partial
x_{3n+i}}=0,\frac{\partial }{\partial t}\left( \frac{\partial L}{\partial
x_{2n+i}}\right) -\frac{\partial L}{\partial x_{7n+i}}=0, \\
\frac{\partial }{\partial t}\left( \frac{\partial L}{\partial x_{3n+i}}%
\right) +\frac{\partial L}{\partial x_{n+i}}=0,\frac{\partial }{\partial t}%
\left( \frac{\partial L}{\partial x_{4n+i}}\right) +\frac{\partial L}{%
\partial x_{6n+i}}=0,\frac{\partial }{\partial t}\left( \frac{\partial L}{%
\partial x_{5n+i}}\right) -\frac{\partial L}{\partial x_{i}}=0, \\
\frac{\partial }{\partial t}\left( \frac{\partial L}{\partial x_{6n+i}}%
\right) -\frac{\partial L}{\partial x_{4n+i}}=0,\frac{\partial }{\partial t}%
\left( \frac{\partial L}{\partial x_{7n+i}}\right) +\frac{\partial L}{%
\partial x_{2n+i}}=0,%
\end{array}
\label{3.13}
\end{equation}%
Thus the equations found in \textbf{Eq. }(\ref{3.13}) are named \textit{%
Euler-Lagrange equations} structured by means of $\Phi _{L}^{J_{5}}$ on
Clifford K\"{a}hler manifold $(M,V)$ and so, the triple $(M,\Phi
_{L}^{J_{5}},\xi )$ is called a \textit{mechanical system }on Clifford K\"{a}%
hler manifold $(M,V)$\textit{.}

Sixth, we present Euler-Lagrange equations for quantum and classical
mechanics by means of $\Phi _{L}^{J_{6}}$ on Clifford K\"{a}hler manifold $%
(M,V).$

Let $J_{6}$ be a local basis on Clifford K\"{a}hler manifold $(M,V).$Let
semispray $\xi $ give as in \textbf{Eq.}(\ref{3.1}). So, \textit{Liouville
vector field} on Clifford K\"{a}hler manifold $(M,V)$ is the vector field
defined by

\begin{equation}
\begin{array}{c}
V_{J_{6}}=J_{6}(\xi )=X^{i}\frac{\partial }{\partial x_{6n+i}}-X^{n+i}\frac{%
\partial }{\partial x_{7n+i}}-X^{2n+i}\frac{\partial }{\partial x_{3n+i}}%
+X^{3n+i}\frac{\partial }{\partial x_{2n+i}} \\
+X^{4n+i}\frac{\partial }{\partial x_{5n+i}}-X^{5n+i}\frac{\partial }{%
\partial x_{4n+i}}-X^{6n+i}\frac{\partial }{\partial x_{i}}+X^{7n+i}\frac{%
\partial }{\partial x_{n+i}}.%
\end{array}
\label{3.14}
\end{equation}%
The function given by $E_{L}^{J_{6}}=V_{J_{6}}(L)-L$ is\textit{\ energy
function} and found by
\begin{equation}
\begin{array}{c}
E_{L}^{J_{6}}=X^{i}\frac{\partial L}{\partial x_{6n+i}}-X^{n+i}\frac{%
\partial L}{\partial x_{7n+i}}-X^{2n+i}\frac{\partial L}{\partial x_{3n+i}}%
+X^{3n+i}\frac{\partial L}{\partial x_{2n+i}} \\
+X^{4n+i}\frac{\partial L}{\partial x_{5n+i}}-X^{5n+i}\frac{\partial L}{%
\partial x_{4n+i}}-X^{6n+i}\frac{\partial L}{\partial x_{i}}+X^{7n+i}\frac{%
\partial L}{\partial x_{n+i}}-L.%
\end{array}
\label{3.15}
\end{equation}%
The function $i_{J_{6}}$ induced by $J_{6}$ and given by%
\begin{equation}
i_{J_{6}}\omega (X_{1},X_{2},...,X_{r})=\sum_{i=1}^{r}\omega
(X_{1},...,J_{6}X_{i},...,X_{r}),  \label{3.16}
\end{equation}%
is said to be \textit{vertical derivation, }where $\omega \in \wedge
^{r}{}M, $ $X_{i}\in \chi (M).$ The \textit{vertical differentiation} $%
d_{J_{6}}$ is determined by%
\begin{equation}
d_{J_{6}}=[i_{J_{6}},d]=i_{J_{6}}d-di_{J_{6}},  \label{3.17}
\end{equation}%
We say the closed K\"{a}hler form is the closed 2-form given by $\Phi
_{L}^{J_{6}}=-dd_{_{J_{6}}}L$ such that

\begin{eqnarray*}
d_{_{J_{6}}} &=&\frac{\partial }{\partial x_{6n+i}}dx_{i}-\frac{\partial }{%
\partial x_{7n+i}}dx_{n+i}-\frac{\partial }{\partial x_{3n+i}}dx_{2n+i}+%
\frac{\partial }{\partial x_{2n+i}}dx_{3n+i} \\
&&+\frac{\partial }{\partial x_{5n+i}}dx_{4n+i}-\frac{\partial }{\partial
x_{4n+i}}dx_{5n+i}-\frac{\partial }{\partial x_{i}}dx_{6n+i}+\frac{\partial
}{\partial x_{n+i}}dx_{7n+i}
\end{eqnarray*}%
and%
\begin{equation}
d_{_{J_{6}}}:\mathcal{F}(M)\rightarrow \wedge ^{1}{}M  \label{3.18}
\end{equation}

Considering \textbf{Eq.} (\ref{1.1}), similar to the above first \ and
second cases , we calculate\ the following equations \
\begin{equation}
\begin{array}{c}
\frac{\partial }{\partial t}\left( \frac{\partial L}{\partial x_{i}}\right) +%
\frac{\partial L}{\partial x_{6n+i}}=0,\frac{\partial }{\partial t}\left(
\frac{\partial L}{\partial x_{n+i}}\right) -\frac{\partial L}{\partial
x_{7n+i}}=0,\frac{\partial }{\partial t}\left( \frac{\partial L}{\partial
x_{2n+i}}\right) -\frac{\partial L}{\partial x_{3n+i}}=0, \\
\frac{\partial }{\partial t}\left( \frac{\partial L}{\partial x_{3n+i}}%
\right) +\frac{\partial L}{\partial x_{2n+i}}=0,\frac{\partial }{\partial t}%
\left( \frac{\partial L}{\partial x_{4n+i}}\right) +\frac{\partial L}{%
\partial x_{5n+i}}=0,\frac{\partial }{\partial t}\left( \frac{\partial L}{%
\partial x_{5n+i}}\right) -\frac{\partial L}{\partial x_{4n+i}}=0, \\
\frac{\partial }{\partial t}\left( \frac{\partial L}{\partial x_{6n+i}}%
\right) -\frac{\partial L}{\partial x_{i}}=0,\frac{\partial }{\partial t}%
\left( \frac{\partial L}{\partial x_{7n+i}}\right) +\frac{\partial L}{%
\partial x_{n+i}}=0,%
\end{array}
\label{3.19}
\end{equation}%
Thus the equations obtained in \textbf{Eq. }(\ref{3.19}) infer \textit{%
Euler-Lagrange equations} structured by means of $\Phi _{L}^{J_{6}}$ on
Clifford K\"{a}hler manifold $(M,V)$ and so, the triple $(M,\Phi
_{L}^{J_{6}},\xi )$ is called a \textit{mechanical system }on Clifford K\"{a}%
hler manifold $(M,V)$\textit{.}

\section{Conclusion}

From above, Lagrangian formalisms has intrinsically been described taking
into account a canonical local basis $\{J_{i}\},$ $i=\overline{1,6}$ of $V$
on Clifford K\"{a}hler manifold $(M,V).$

The paths of semispray $\xi $ on Clifford K\"{a}hler manifold are the
solutions Euler--Lagrange equations raised in (\ref{3.6}), (\ref{3.13}) and (%
\ref{3.19}), and also obtained by a canonical local basis $\{J_{i}\},$ $i=%
\overline{1,6}$ of vector bundle $V$ on Clifford K\"{a}hler manifold $(M,V)$%
. \ One may be shown that these equations are very important to explain the
rotational spatial mechanics problems.

\end{document}